\documentclass{aa} 

\usepackage{graphics} 
\usepackage{astron} 
\usepackage{times} 

\newcommand{\kms}{km~s$^{-1}$}
\newcommand{\kmskpc}{km~s$^{-1}$~kpc$^{-1}$}
\newcommand{\s}{\sigma} 
\newcommand{\ci}{_\circ}

\begin{document}

%-------%
% Title %
%-------%
\title{On the Evolution of Moving Groups: An Application to the Pleiades Moving
Group}

\titlerunning{On the Evolution of Moving Groups}

\author{ R. Asiain \and F. Figueras \and J. Torra }

\thesaurus{08.11.1;10.11.1;10.19.1} 

\offprints{R. Asiain} \mail{rasiain@am.ub.es} 

\institute{Departament  d'Astronomia i Meteorologia,  Universitat de
           Barcelona, Avda. Diagonal 647, E08028, Barcelona, Spain }

\date{Received ; Accepted} 

\maketitle 

%----------%
% Abstract %
%----------%

\begin{abstract}

 The disruption of stellar systems, such as open clusters or stellar complexes, 
stands out as one of the most reasonable physical processes accounting for the 
young moving groups observed in the solar neighbourhood. In the present 
study we analyse some of the mechanisms that are important in the kinematic 
evolution of a group of unbound stars, such as the focusing phenomenon and 
its ability to recover the observed moving group's velocity dispersions, and 
the efficiency of disc heating and galactic differential rotation in 
disrupting unbound stellar systems. Our main tools used to perform this analysis are both the epicycle 
theory and the integration of the equations of motion using a realistic 
gravitational potential of the Galaxy.

  The study of the trajectories followed by stars in each of the Pleiades 
moving group substructures found by Asiain et al.~\cite*{asiain2} allows us to 
determine their stellar spatial and velocity distribution evolution. The
kinematic properties of these substructures are compared to those of a
simulated stellar complex which has evolved under the influence of the
galactic gravitational potential and the disc heating. We conclude that a
constant diffusion coefficient compatible with the observational heating law 
is able to explain the velocity and spatial dispersions of the Pleiades moving 
group substructures that are younger than $\sim 1.5 \cdot 10^8$~yr.


\end{abstract}

%--------------% 
% Introduction %
%--------------% 

\section{Introduction} \label{sintro}

 A ``supercluster" can be defined as a group of stars gravitationally
unbound that share the same kinematics and may occupy
extended regions of the Galaxy. A ``moving group"
(MG hereafter) or ``moving cluster", however, is the part of the supercluster 
that can be observed from the Earth (see, for instance, Eggen 1994).

 Independent of the physical process that leads to the formation of 
superclusters, there are basically two factors acting against their persistence 
in the general stellar background. First, when a group of unbound stars 
concentrated in the phase space has a certain velocity 
dispersion, the galactic differential rotation tends to spread them out very 
quickly in the direction of the galactic rotation (e.g., Woolley 1960). Second, 
observations show that the velocity dispersion of disc stars is increased with 
their age, which is usually interpreted as the result of a continuous process 
of gravitational acceleration that may be produced by different agents (e.g.,
Lacey 1991).  This second factor, generally referred to as {\em disc heating}, 
also disperses stars very efficiently. It is therefore striking to verify that 
some of the classical MGs are several $10^8$~yr old. Some authors have overcome 
part of this problem by 
assuming that the velocity dispersion of these stellar groups must be very 
small -- e.g.  Eggen \cite*{eggen89}, Yuan \& Waxman \cite*{yuanwax}, 
Soderblom \& Mayor \cite*{sodermay}. Nevertheless, some recent studies 
\cite{chen,sabas,cesca1,chereul2,asiain2} do not support such a small 
dispersions.

  There have been several attempts to explain the origin of superclusters.
Perhaps one of the most widely accepted explanation is the ``evaporation" of 
the outermost stellar component (or corona) of open clusters (e.g. Efremov 
1988). Over time, open clusters are disrupted by the gravitational interaction 
with massive objects in the Galaxy (such as giant molecular clouds), and as a
result the open cluster members fill a long tube 
in space. The part of this tube that we can observe should have a small 
velocity dispersion -- e.g. Weidemann et al. \cite*{weide}, Eggen 
\cite*{eggen94}.  Although based in old stellar evolutionary models by 
Iben~\cite*{iben}, Wielen~\cite*{wielen71} found that about 50\% of the open 
clusters disintegrate in less than 2$\cdot$10$^8$ years, a result supported by
posterior studies \cite{terle87,theuns92}. If superclusters come from the 
evaporation of open clusters's coronae, then Wielen's~\cite*{wielen71} results 
indicate that most superclusters should be young or the last tracers of former 
clusters. 

MGs may also be produced by the dissolution of larger stellar agglomerations,
such as stellar complexes or fragments of old spiral arms 
\cite{woolley,wielen71,yuan,elme85,efre88,come97}. Thus, a MG could be a 
mixture of stars coming from different open clusters' coronae and disrupted 
associations that share the same origin (and motion). This process of MG 
formation includes the first hypothesis mentioned here. Alternatively, Casertano et al.~\cite*{caser93} proposed that MGs 
are open clusters that evolved under the gravitational influence of a 
large local mass that surrounds and traps them. However, it seems difficult
to specify the nature of such a large mass. Finally, it has been proposed 
that superclusters could actually be made of stars trapped around stable 
periodic orbits (e.g. M\"ullari et al. 1994; Raboud et al. 1998).

 In this paper we assume that superclusters occupied a small volume in the 
phase space in their first stages, when they became gravitationally unbound.
From that point on, they evolved under the influence of the 
gravitational potential of the Galaxy, as it would be the case in the classical 
picture of stellar evaporation from open clusters' coronae. We then analyse 
different aspects related to the evolution of MGs.  

First, in Sect.~\ref{sfoc} we use epicycle approximation to study the ability of 
the ``focusing phenomenon" \cite{yuan} to periodically group MG stars when
no disc heating is considered. In Sect.~\ref{sevol} we show how to determine 
the stellar trajectories from an analytic expression of the galactic potential.
The disc heating effect on moving groups' properties can be considered by 
introducing random perturbations to the velocity components of stars. These
trajectories allow us to study the evolution of unbound stellar systems
independent of the local galactic properties (in contrast with epicycle
approximation). 

 We focus our study on the origin and evolution of the Pleiades moving group, since 
it is the youngest moving group in the solar neighbourhood and, therefore, we 
may find it easier to recover its past properties with certain confidence 
(Sect.~\ref{smgtraj}). We consider the different Pleiades substructures found 
in Asiain et al.~(1999, Paper~I), and estimate their 
kinematic age and galactic position at birth. Finally, in Sect.~\ref{ssevsc} 
we simulate a stellar complex and determine the trajectory of its stars to study
its disruption over time. The disc heating effect on the evolution of the
stellar complex is evaluated.   

\section{The focusing phenomenon}\label{sfoc}

 The epicycle approximation allows us to study, in an analytical way, the 
evolution of an unbound system of stars under the influence of the galactic
potential. Under such approximation,  the 
equations of motion of stars can be expressed as:
\begin{eqnarray}\label{epicm}
\xi^\prime & = & \xi^\prime_a + \xi^\prime_b \cos(\kappa t + \phi) \nonumber \\
\eta^\prime & = & \eta^\prime_a - 2 \mbox{A} \xi^\prime_a t - 
   \displaystyle\frac{2 \omega\ci \xi^\prime_b}{\kappa} \sin(\kappa t + \phi) 
	\nonumber \\
\zeta^\prime & = & \zeta^\prime_a \cos(\nu t + \psi)
\end{eqnarray}
in the coordinate system ($\xi^\prime,\eta^\prime,\zeta^\prime$) centered 
at the current position of the Sun.  $\xi^\prime$ points towards the  
galactic center (GC), $\eta^\prime$ is a linear 
coordinate measured along a circumference of radius R$\ci$ (galactocentric
distance of the Sun) and positive in the sense of the galactic rotation
(GR), and $\zeta^\prime$ points towards the north galactic pole (NGP).
$\kappa$ is the {\em epicyclic frequency}, $\nu$ is the {\em vertical 
frequency}, $t$ is the time (=0 at present) and $\omega\ci$ is the angular
velocity of the Galaxy at the current position of the Sun. 
$\xi^\prime_a , \xi^\prime_b, \eta^\prime_a,
\zeta^\prime_a, \phi$ and $\psi$ are integration constants related to the 
current position
and velocity of a star ($\xi^\prime\ci$, $\eta^\prime\ci$, $\zeta^\prime\ci$, 
$\dot{\xi}^\prime\ci$, $\dot{\eta}^\prime\ci$, $\dot{\zeta}^\prime\ci$) as: 
\begin{eqnarray}\label{epic2}
\xi^\prime_a & = & - \displaystyle\frac{2 \xi^\prime\ci \omega\ci + \dot{\eta}^\prime\ci}{2 \mbox{B}} \vspace{.15cm}  \nonumber \\
\xi^\prime_b & = & - \left[ \left( \displaystyle\frac{2 \mbox{A} \xi^\prime\ci + \dot{\eta}^\prime\ci }{2 \mbox{B}} \right)^2
            + \left( \displaystyle\frac{\dot{\xi}^\prime\ci}{\kappa} \right)^2 \right]^{1/2} \vspace{.15cm} \nonumber  \\
\eta^\prime_a & = & \eta^\prime\ci - \displaystyle\frac{2 \omega\ci \dot{\xi}^\prime\ci}{\kappa^2} \vspace{.15cm}\nonumber   \\
\zeta^{\prime 2}_a & = &  \zeta^{\prime 2}\ci + \displaystyle\frac{\dot{\zeta}^{\prime 2}\ci}{\nu^2} \vspace{.15cm} \nonumber  \\
\phi & = & \arctan \left( \displaystyle\frac{ - 2 \dot{\xi}^\prime\ci \mbox{B}}{\kappa
          (\dot{\eta}^\prime\ci + 2 \xi^\prime\ci \mbox{A})} \right) \vspace{.15cm} \nonumber  \\
\psi & = & \arctan \left( \displaystyle\frac{- \dot{\zeta}^\prime\ci}{\nu \zeta^\prime\ci} \right)
\mbox{ ,}
\end{eqnarray}
where the B and A are the Oort's constants. In \ref{aerrorep} 
we use Eqs.~\ref{epicm} to evaluate the evolution of space and velocity 
dispersions of a group of stars with low peculiar velocity with respect
to their {\em Regional Standard of Rest} (RSR). Eqs.~\ref{epic5} show that 
dispersions in position and velocity components oscillate around constant
values, except for the azimuthal coordinate $\eta^\prime$, whose dispersion
$\s_{\eta^\prime}$ increases with $t$. After a few $10^7$~yr, this increase
is dominated by the secular terms for typical position and velocity dispersions
of young stellar groups, and it can be approximated by the linear 
relationship
\begin{equation}\label{erecta}
\s_{\eta^\prime} = \displaystyle\frac{\mbox{A}}{\mbox{B}}
\left( 4 \omega\ci^2 \s_{\xi^\prime\ci}^2 + \s_{\dot{\eta}^\prime\ci}^2 \right)^{1/2} t \; \mbox{ . }
\end{equation}
 If we consider a sample of stars with $\s_{\xi^\prime\ci}\approx$~0~pc and 
$\s_{\dot{\eta}^\prime\ci} \approx$~0~\kms, then all dispersions in position 
and velocity would oscillate around their mean values with an epicycle frequency
$\kappa$. Stars would meet every $\Delta t = \displaystyle\frac{2 \pi}{\kappa}$
(at the position of the Sun $\Delta t \approx 1.5 \cdot 10^8$~yr), a fact that
has been referred to as ``focusing phenomenon" \cite{yuan}. However, when we 
apply Eq.~\ref{erecta} to a more realistic case where dispersions are close to 
those of the open clusters and associations, i.e. a few parsecs in 
$\s_{\xi^\prime\ci}$ and $\sim$~1-2~\kms\ in $\s_{\dot{\eta}^\prime\ci}$, we 
obtain $\s_{\eta^\prime}\approx$ 1~kpc after a $t \approx$~5-10~$\cdot 10^8$~yr
-- a period of time shorter than the age of some moving groups detected in the 
solar neighbourhood (e.g. Chen et al. 1997; Paper~I). Thus, galactic 
differential rotation disrupts very efficiently unbound systems of stars. 
Nonetheless, if the distribution of stars in $\eta^\prime$ is supposed to be 
gaussian after the disruption of the stellar system, an important proportion of 
the original sample will be still concentrated in space after a long period of 
time; for instance, for $\s_{\eta^\prime\ci}$~=~1~kpc we can find still 
$\sim$~24~\% of the initial stars in a region 600~pc long in $\eta^\prime\ci$ 
after 5-10~$\cdot 10^8$~yr.

\section{Stellar trajectories}\label{sevol}

 The epicycle approximation holds for stars that follow almost circular orbits 
around the center of the Galaxy. Although valid for most of our stars, some
of them, with high peculiar velocities with respect to the {\em Local Standard of 
Rest} (LSR), perform radial galactic excursions of a few kpc in length. The 
galactic gravitational potential changes significatively at different points
of their trajectories, and so the first order approximation is no longer valid. 
In addition, the vertical motion is poorly described by a harmonic 
oscillation as stars gain certain height over the galactic plane. For a 
more rigorous analysis, we need to use a realistic model of the galactic 
gravitational potential. Stellar orbits will be determined from this model by 
integrating the equations of motion (Sect.~\ref{ssstetraj}). Moreover,
the addition of a constant scattering in this process will allow us to  
account for the disc heating effect on the stellar trajectories (\ref{ssdh}).

\subsection{The galactic potential}
\label{ssstetraj}

  Expressed in a cartesian coordinate system ($\xi, \eta, \zeta$) centered
at the position of the Sun\footnote{$\xi$ points towards the GC,
$\eta$ towards the sense of the GR, and $\zeta$ towards the NGP} and rotating 
at a constant angular velocity $\omega\ci$,
\begin{equation}\label{epot}
 \omega\ci = \sqrt{\displaystyle\frac{1}{ R\ci} \;\;
\Biggl( \frac{\partial \Phi}{\partial R} \Biggr)\ci} \; \; \mbox{  ,}
\end{equation} 
the equations of motion of a star, assuming that the gravitational potential 
of the Galaxy $\Phi_\mathrm{G}(R,\theta,z;t)$ (in cylindrical coordinates 
centered on the GC) is known, are:
\begin{eqnarray} \label{emot1}
{\ddot \xi} & = &  - \displaystyle\frac{\partial \Phi_\mathrm{G}}{\partial \xi} 
- \omega\ci^2 (R\ci - \xi) - 2 \omega\ci \dot{\eta} \nonumber \\
{\ddot \eta} & = & - \displaystyle\frac{\partial \Phi_\mathrm{G}}{\partial \eta}
               + \omega\ci^2 \eta + 2 \omega\ci \dot{\xi} \nonumber  \\
{\ddot \zeta} & = & 
               - \displaystyle\frac{\partial \Phi_\mathrm{G}}{\partial \zeta}  
\; \; \mbox{ .}
\end{eqnarray}
A fourth order Runge-Kutta integrator allows us to numerically solve these
equations when $\Phi_\mathrm{G}$ is known, obtaining the trajectory
of the star. To get a realistic estimation of the galactic gravitational 
potential we decompose it into three parts: the general axisymmetric 
potential $\Phi_\mathrm{AS}$, the spiral arm $\Phi_\mathrm{Sp}$, and the 
central bar $\Phi_\mathrm{B}$ perturbations to the first contribution, i.e.
\begin{eqnarray*}
\Phi_\mathrm{G} = \Phi_\mathrm{AS} +  \Phi_\mathrm{Sp} + \Phi_\mathrm{B} 
\mbox{ .}
\end{eqnarray*}

We adopt the model developed by Allen \& Santill\'an \cite*{allen91} for 
the axisymmetric part of the potential, $\Phi_\mathrm{AS}(R,z)$, because of both 
its mathematical simplicity -- which allows us to determine orbits with a very 
low CPU consumption -- and its updated parameters. The model consists of a 
spherical central bulge and a disk, both of the Miyamoto-Nagai~\cite*{miya75} 
form, plus a massive spherical halo. This model is symmetrical with respect to an 
axis and a plane.  The authors adopted the recommendations of the IAU 
\cite{kerr} for the galactocentric distance of the Sun ($R\ci$ = 8.5~kpc) and
the circular velocity at the position of the Sun ($\Theta\ci$=~220~\kms).
The Oort constants derived from this model are well within the currently 
accepted values.

 Spiral arm perturbation to the potential is taken from Lin and associates'
theory (e.g. Lin 1971 and references therein), that is:
\begin{equation}\label{ebrac}
\Phi_\mathrm{Sp}(\mbox{R},\theta;t) = {\cal A} \cos ( m (\Omega_p t - \theta) + 
\phi(R) ) \; \; \mbox{ ,}
\end{equation}
where 
\begin{eqnarray*}
 {\cal A} & = & \displaystyle\frac{ (R\ci \omega\ci)^2 f_{r0} \tan i}{m} 
 \mbox{ ,}\\ 
 \phi(R) & = & - \displaystyle\frac{m}{\tan i} \ln \left(  
\displaystyle\frac{R}{R\ci} \right)  + \phi_0 \mbox{ .}
\end{eqnarray*}
${\cal A}$ is the amplitude of the potential, $f_{r0}$ is the ratio between
the radial component of the force due to the spiral arms and that due to the 
general galactic field.
$\Omega_p$ is the constant angular velocity of the spiral pattern,
$m$ is the number of arms, $i$ is the pitch angle, $\phi$ is the radial phase of
the wave and $\phi_0$ is a constant that fixes the position of the minimum
of the spiral potential. According to Yuan~\cite*{yuan69}, we adopt  
$f_{r0} = 0.05$ (this value has been confirmed in a recent study by 
Fern\'andez, 1998) and $\Omega_p = 13.5$~\kmskpc.

We consider a classical two arm pattern ($m = 2$) \cite{yuan69,vallee95}.
If we assume that the Sagittarius arm is located at a galactocentric distance 
R$_{Sag} =$~7.0~kpc, and that the interarm distance in the Sun position is 
$\Delta R = $3.5~kpc as suggested by observations on spiral arm tracers 
\cite{becker70,geor76,liszt85,kurtz94}, then the pitch angle can be 
determined to be:
\begin{equation}\label{einterarm}
\tan i = \displaystyle\frac{m}{2 \pi} \ln ( 1 + R_{Sag}^{-1} \; \Delta R )
\mbox{ .}
\end{equation}
For $m = 2$, we obtain $i = 7\fdg35$ (close to the Yuan's (1969) value
$i = 6\fdg2$).

 For $t=0$ and $\theta = 0\degr$, the adopted value $\phi_0$ leads to a 
minimum in the potential at the observed position of the Sagittarius arm 
(R = R$_{Sag}~=~7.0$~kpc, $l = 0\degr$). $\phi_0$ can thus be expressed as:
\begin{equation}\label{ephi0}
\phi_0  = \pi + \displaystyle\frac{m}{\tan i} \ln \left(
\displaystyle\frac{R_{Sag}}{R\ci} \right)  .
\end{equation}
For $i= 7\fdg 35$ and $m=2$ we obtain $\phi_0  =$~0.131~rad.

The central bar potential we use here is, for simplicity, a triaxial ellipsoid 
with parameters taken from Palou\u s et al. \cite*{palous93}, i.e.
\begin{equation}
\Phi_b(\mbox{R},\theta,z;t) = - \displaystyle\frac{ G M_{bar}}{\left( q_{bar}^2
+ x^2 + \displaystyle\frac{a_{bar}^2}{b_{bar}^2} y^2 +
\displaystyle\frac{a_{bar}^2}{c_{bar}^2} z^2  \right)^{1/2}},
\end{equation}
where $x = \mbox{R}\ci \cos (\theta - \Omega_B t - \theta\ci )$ and 
$y = \mbox{R}\ci \sin (\theta - \Omega_B t - \theta\ci) $, with $\theta\ci =$~45$\degr$
\cite{white92}.   $a_{bar}, b_{bar}$ and
$c_{bar}$ are the three semi-axes of the bar, with $q_{bar}$ its scale
length, and with $\displaystyle\frac{a_{bar}}{b_{bar}} =  
\displaystyle\frac{1}{0.42}$,
$\displaystyle\frac{a_{bar}}{c_{bar}} =  \displaystyle\frac{1}{0.33}$,
and $q_{bar} = 5$~kpc. The adopted total mass of the bar, $M_{bar}$, is 
$10^9$~M$_\odot$, and its angular velocity, $\Omega_B$, is 70~\kmskpc\ 
\cite{binney91}. Although most of the parameters that define the bar are 
very uncertain, the effect of the bar on the stellar trajectories becomes
important only after several galactic rotations, which requires a length of
time greater than the age of the stars considered in our study.

\subsection{Disc heating in the stellar trajectories} \label{ssdh}

 The observational increase in the total stellar velocity dispersion ($\s$) with 
time ($t$), or {\em disc heating}, can be approximated by an equation of the form:
\begin{equation}\label{eheat0}
\s(t)^n = \s\ci^n  + C_v t \mbox{ ,}
\end{equation}
where $\s\ci$ is the dispersion at birth and $C_v$ the ``apparent diffusion 
coefficient" \cite{wielen77}. The constants $n$, $\s\ci$ and $C_v$ give one 
important information on the physical mechanism 
responsible for the disc heating \cite{lacey91,fridman94}.

\begin{figure}
      \mbox{}
      \vspace{6.5cm}
      \includegraphics{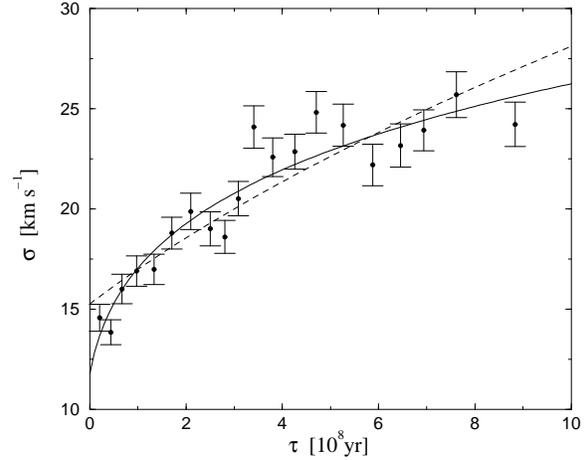}
      \caption[]{Fit of Equation~\ref{eheat0} to our data (see text), using 
      		 stars closer than 300~pc from the Sun, with 100 stars per bin.
                 {\em Solid line}: $\s\ci$, $C_v$ and $n$ are fitted;                
                 {\em dashed line}: $\s\ci$ and $C_v$ are fitted, whereas
                 $n$ is fixed to 2. The error bars on points include only
                 statistical uncertainties, calculated as $\epsilon_\s =
                 \s / \sqrt{2 N}$, where $N$ is the number of stars per bin}  
      \label{fheac0}
    \end{figure}

 Our sample of B and A main sequence type stars, described in
Paper~I, allows us to determine an accurate and detailed 
disc heating law for the last $10^9$~yr, given the quality and uniformity of 
our ages and the size of this sample (2\,061 stars). A standard nonlinear 
least-squares method ({\em Levenberg-Marquardt} method) has been applied to fit 
the heating coefficients to our data (Fig.~\ref{fheac0}). In this way we 
obtain $\s\ci \approx 12$~\kms, $n \approx 5$ and $C_v \approx
0.01$~(\kms)$^n$yr$^{-1}$, similar to Lacey's~\cite*{lacey91} results for 
$n = 5$, i.e. $\s\ci = 15$~\kms\ and $C_v = 0.01$~(\kms)$^5$yr$^{-1}$. When we 
fix $n$ to 2, we obtain $\s\ci \approx 15$~\kms  and 
$C_v \approx 5.6 \cdot 10^{-7}$~(\kms)$^2$yr$^{-1}$, which is again in 
agreement with Lacey's~\cite*{lacey91} 
($\s\ci = 15$~\kms\ and $C_v = 5 \cdot 10^{-7}$~(\kms)$^2$yr$^{-1}$) and 
Wielen's~\cite*{wielen77} ($\s\ci = 10$~\kms\ and 
$C_v = 6 \cdot 10^{-7}$~(\kms)$^2$yr$^{-1}$) fits with $n = 2$. 

Our accurate observational heating law (circles in Fig.~\ref{fheac0}) is far 
from being smooth. Two special features can be observed on this law: first, the
velocity dispersion shows a steep increase during the first 
$\sim 4 \cdot 10^8$~yr, probably due to the phase mixing of young stars.
After this point, this increase becomes less pronounced. Second, an almost
periodic oscillation seems to be superimposed over the ``continuous"
heating law. This oscillation (period $\approx 3\cdot 10^8$~yr) could be the 
signature of an episodic event (see for instance Binney \& Lacey 1988, Sellwood
1999). 
   
   \begin{figure}
      \mbox{}
      \vspace{6.5cm}
      \includegraphics{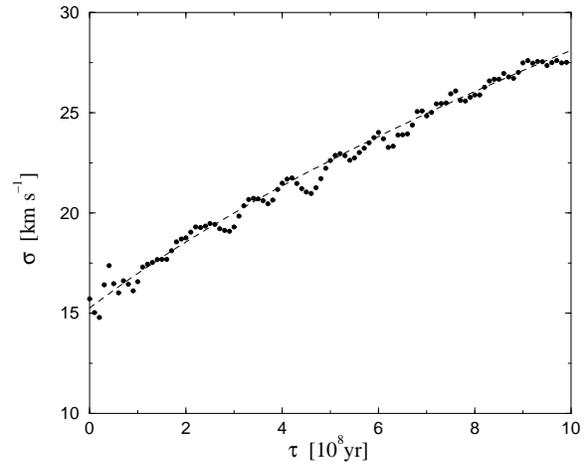}
      \caption[]{Velocity dispersion vs. time, determined by integrating
                 the equations of motion of a simulated sample when an
                 isotropic diffusion is applied at each time-step.
                 {\em Dashed line}: observational
                 fit to disc heating with a fixed slope $n$=2}
      \label{fheac}
    \end{figure}

Since we do not know the details about the mechanisms that perturb the 
stellar orbits and cause the disc heating effect, we assume for 
simplicity that the accelerating processes acting on a given star can be 
approximated by a sequence of independent and random perturbations of short 
duration \cite{wielen77}. In order to evaluate the magnitude
of these perturbations we generated a sample of 500 stars located around the 
position of the Sun at $t = 0$, and uniformly distributed in a 100~pc/side 
cube. Initially, these stars are moving with the same velocity as the LSR, with 
an isotropic dispersion in velocity $\s\ci$. During the process of orbit 
determination, the velocity of the star is instantaneously changed 
by $\Delta \vec{v}$ every $\Delta$~yr. This $\Delta \vec{v}$ 
is normally distributed around zero with a (constant) isotropic dispersion 
$\sigma_h$ in each component. The equations of motion are integrated using 
only the axisymmetric part of the galactic potential ($\Phi_\mathrm{AS}$). 
This allows us to introduce the heating effect only through the parameter 
$\sigma_h$, avoiding possible redundances, such as the scattering of disc stars
by spiral arms. The best approximation to the 
observational heating law is obtained when $\sigma_h = 1.45$~\kms\ for 
$\Delta t = 10^7$~yr, and 
$\s\ci \approx$~15~\kms. It can be demonstrated that $\sigma_h$ and $\Delta t$
are related with the ``true diffusion coefficient" $D$ introduced by
Wielen~\cite*{wielen77} just as $D=\displaystyle\frac{\sigma_h^2}{\Delta t}$. 
On the other hand $C_v = 2.95 \; D$, as demonstrated by Wielen~\cite*{wielen77} 
using the epicycle theory. We obtain in this way an independent estimation of 
$C_v$, i.e. $C_v= 2.95 \; \displaystyle\frac{1.45^2}{10^7} = 6.2 
\cdot 10^{-7}$~(\kms)$^2$yr$^{-1}$. This value is in an excellent  
agreement with our previous estimation. In Figure~\ref{fheac} the 
evolution of the total velocity dispersion for this simulated sample is compared 
with the fit of Eq.~\ref{eheat0} to the observational data when $n = 2$, showing 
an excellent match. This procedure will be used in Sect.~\ref{ssevsc} to 
simulate the evolution of a stellar system under the influence of the disc 
heating.

\section{The Pleiades Moving Groups}\label{smgtraj}

 In Paper~I we used a method based on 
non-parametric 
density estimators to detect MGs among a sample of 2\,061 B and A main sequence
type stars in the four-dimensional space (U,V,W,log (age) ). We used 
HIPPARCOS data as well as radial velocities from several sources (see Paper~I 
for the details) to determine the stellar spatial velocities, and 
Str\"omgren photometry for the stellar ages. Tables 1 and 2 in Paper~I show 
the main properties of the MGs when separated in the (U,V,W,log (age)) and 
(U,V,log (age)) 
spaces respectively. Since the W-velocity component is less discriminant
than the other three variables, we will focus our study on the MGs 
in Table~2 of Paper~I (Table~\ref{tmgfi1} here). In particular, as already
mentioned in the Sect.~\ref{sintro}, we will deal with those MGS whose 
velocity components resemble those of Pleiades open cluster.

\begin{table}[htb]
\caption[]{Number of members ($\cal N$), velocity components and ages (along 
with their dispersions) of the Pleiades MGs found in Paper~I}\label{tmgfi1}
\begin{center}
\begin{tabular}{cccccc}
\hline
Moving & $\cal N$        & U               &  V               &  W               & $\tau$ \\
 Group &        & [km s$^{-1}$]   &  [km s$^{-1}$]   &  [km s$^{-1}$]   & [10$^8$~yr] \\
\hline
 B1 &   34            & -4.5$_{(4.7)}$ &  -20.1$_{(3.3)}$   &  -5.5$_{(1.9)}$    &   0.2$_{(0.1)}$ \\
 B2 &   75            & -10.7$_{(5.3)}$ &  -18.8$_{(3.7)}$   &  -5.6$_{(2.2)}$    &   0.6$_{(0.2)}$\\
 B3 &   50            & -16.8$_{(5.1)}$ &  -21.7$_{(2.7)}$   &  -5.6$_{(4.6)}$    &   3.0$_{(1.2)}$  \\
 B4 &   53            & -8.7$_{(4.8)}$ &  -26.4$_{(3.3)}$   &  -8.5$_{(4.7)}$    &   1.5$_{(0.5)}$       \\
\hline
\end{tabular}
\end{center}
\end{table}

It is interesting to compare these results
with those recently obtained by Chereul et al. \cite*{chereul2}, who also 
found similar substructures in the Pleiades and other MGs by means of a wavelet 
analysis performed at different scales. The velocity dispersions of their 
substructures are quite a bit larger than those found for classical MGs, in 
agreement 
with our results. In their analysis, they did not consider the stellar 
age as a discriminant variable, which prevents them from detecting those
substructures that are strongly defined in age but not so well defined in the velocity 
components. Another important difference between both methods is that 
Chereul et al. \cite*{chereul2} did not determine photometric ages for A0 to A3 
stars, since no reliable metallicity is available for them. Instead, they 
computed a ``paliative" age that produces an artificial peak in age of 
$10^7$~yr, and a lack of other young stars (up to $\sim 5 \cdot 10^8$~yr). As 
mentioned in Asiain et al.~\cite*{asiain0}, a metallicity $Z = 0.02$ is 
representative, in a statistical sense, of (normal) A type stars. Using this 
value, we did not observe any lack of stars in the young part of the age 
distribution (Paper~I). On the other hand, since we do not use F type stars in 
our study because of the high uncertainties involved in the process to 
determine their ages, our data do not allow us to confirm the existence of the 
$\sim 10^9$~yr old Pleiades substructure found by Chereul et al. \cite*{chereul2}.

\begin{figure}
      \mbox{}
      \vspace{7.2cm}
      \includegraphics{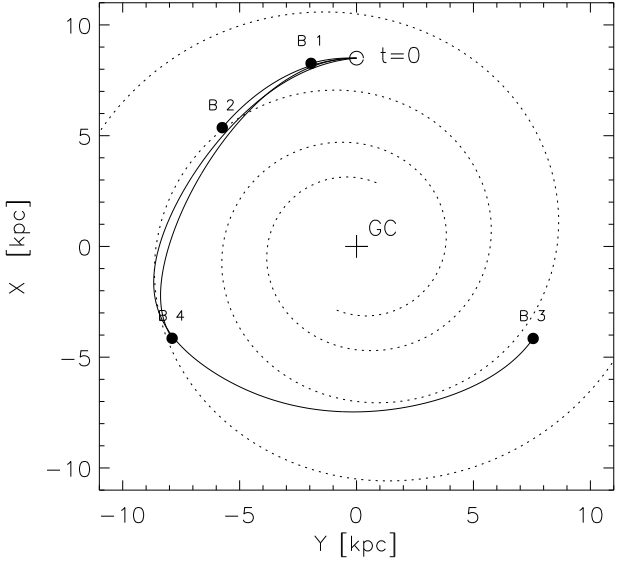}
      \caption[]{Trajectories of the MG's nuclei in Table~\ref{tmgfi1} 
		 backwards in time, from present ($t=0$) until their mean age. 
		 The dashed lines represent spiral arms, as defined in 
		 Sect.~\ref{ssstetraj}. The reference system is rotating 
		 with the same angular velocity as the spiral arms ($\Omega_p$)}
      \label{ftraj}
    \end{figure}

 Using the numerical integration procedure described in Sect.~\ref{ssstetraj}, 
and the mean properties (nuclei) of the Pleiades substructures found in Paper~I 
(Table~\ref{tmgfi1}), we have computed the trajectories of these substructures 
from the present up to the moment they were born (Fig.~\ref{ftraj}). This 
latter age is defined as the average age of the MGs constituent members. 
The youngest group, i.e. B1, is composed of Scorpio-Centaurus (Sco-Cen) OB 
association members (Paper~I), and it was born in the interarm region. In 
Sect.~\ref{ssscen} we study the evolution of this group. Since B1 is still 
too young to be affected by the disc heating effect or phase space mixing,
we can determine its kinematic age with some confidence. The B2 group is 
considered separately in Sect.~\ref{ssb2}. This is also quite a young group and 
contains accurate information on some of the closest associations. The 
birthplace of the older groups, i.e. B3 and B4, is close to a minimum of the spiral arm 
potential, which seems consistent with their being born around this structure. 
Details on their spatial and velocity evolution are given in 
Section~\ref{ssolder}. 

 In recent studies based on the velocity field of Cepheids, Mishurov et 
al.~\cite*{mishu97} and Mishurov \& Zenina~\cite*{mishu99} obtained a set of 
spiral arm parameters which clearly differ from those adopted here (e.g., in 
Mishurov \& Zenina~\cite*{mishu99}, $\phi_0 = 142 \degr$ and $\Omega_p - 
\omega\ci \approx 0.5$~\kmskpc). When considering these new parameters, all 
Pleiades MG substructures turn out to be also placed, at birth, around the  
spiral arms. In this particular case, the Pleiades substructures appear, at
birth, concentrated in a small area of the Galaxy. We also tested that
the main conclusions of the current paper does not critically change
with the selection of any of this two spiral arm patterns, as expected.

\subsection{Scorpio-Centaurus association}\label{ssscen}

\begin{figure}
     \mbox{}
     \vspace{14.cm}
     \includegraphics{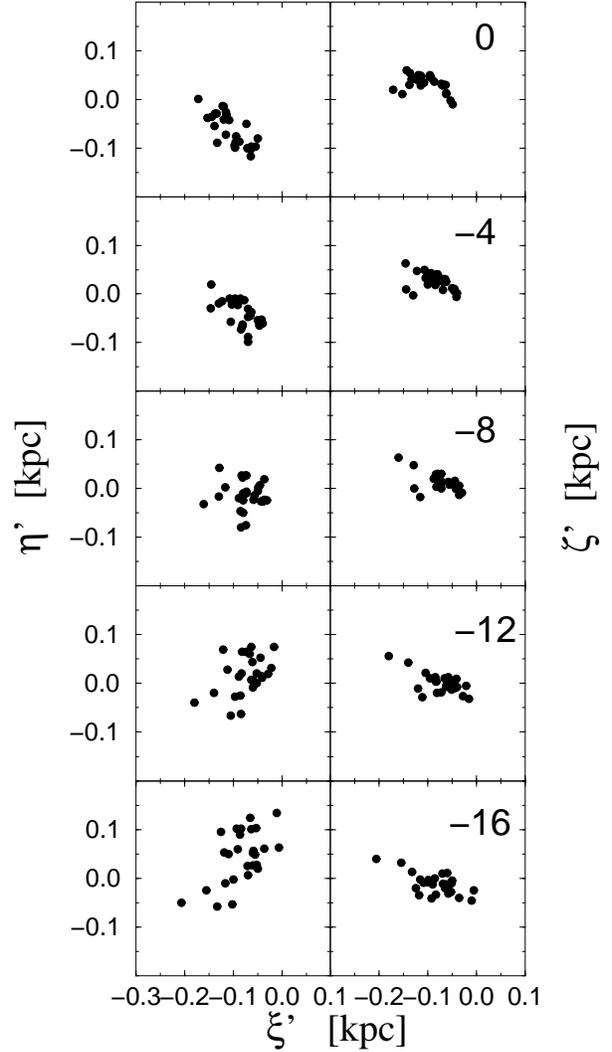}
     \caption[]{Spatial distribution of the 27 stars in B1 sub-group that
                are spatially concentrated in space as a function of time. 
		Numbers in the upper right corners indicate the time in Myr}
     \label{fscoc}
   \end{figure}

 Because of the short age of these stars and the quality of our data we are able
to determine quite precisely these stars' kinematic age, defined as the time at which
they were most concentrated in space -- assuming that they are
gravitationally unbound. We consider here only those stars in B1 that are 
concentrated in space (see Fig.7 in Paper~I). With the exception of
HIP84970 and HIP74449, all of these stars were classified as members of 
Sco-Cen association by de~Zeeuw et al. \cite{dezeeuw99}
\footnote{Though B1 is considered as a single group because of the few members 
it contains, most of its stars belong to three neighbouring associations: 
Upper Scorpius (US), Upper Centaurus Lupus (UCL), and Lower Centaurus Crux 
(LCC) \cite{degeus89,blaauw91,dezeeuw99}.}.
Their position in the galactic and meridional
planes as a function of time are shown in Figure~\ref{fscoc}. It is quite
evident that between around 4 and 12 Myr ago these stars were closer to each 
other than they are at present. However, observational errors
produce additional dispersions on velocity and position, and therefore the
maximum spatial concentration we find is shifted to the present. 
Figure ~\ref{fscoc} also shows us that the last intersection of the 
Sco-Cen association with the galactic plane was between 8 and 16 million years 
ago.

 The evolution of the errors or dispersions in the stellar position and 
velocity can be calculated more precisely by means of the epicycle approximation
(Eqs.~\ref{epic5}). We now consider the fact that the observed dispersion 
in position 
inside a given MG ($\s_{{\rho} , {\mathrm obs}}$, where $\rho$ is the distance 
from the LSR to a given star) can be decomposed into two parts, i.e.
\begin{equation}
\s^2_{{\rho} , {\mathrm obs}} (t) = \s^2_{{\rho} , {\mathrm int}}(t)+ 
	\s^2_{{\rho} , {\mathrm err}} (t)\mbox{ ,}
\end{equation}
where

\noindent
$\s_{{\rho} , {\mathrm int}} (t)$ is the MG intrinsic dispersion at the
time $t$, and \\
$\s_{{\rho} , {\mathrm err}} (t)$ is the MG mean observational error.\vspace{.1cm}

\begin{figure}
      \mbox{}
      \vspace{6cm}
      \includegraphics{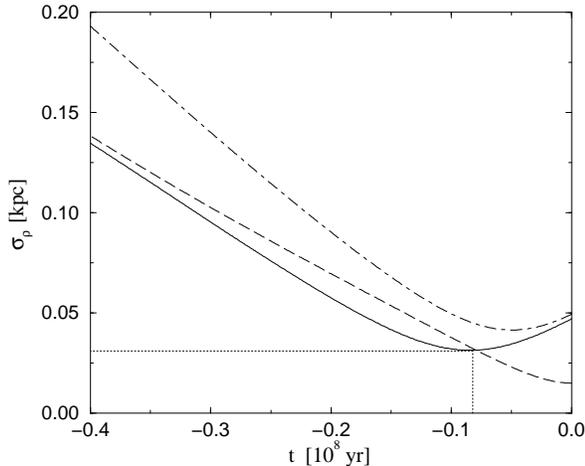}
      \caption[]{Evolution of the observed dispersion ({\em
           dot-dashed line}), mean propagated error ({\em dashed line})
           and intrinsic dispersion ({\em solid darker line})
           in $\rho$ (distance to the Sun), determined from the most spatially 
	   concentrated stars in the B1 group}
      \label{fscoc1}
    \end{figure}

\noindent
Since both $\s_{{\rho} , {\mathrm obs}}$ and $\s_{{\rho} , {\mathrm err}}$ can 
be easily determined at different epochs ($\s_{{\rho} , {\mathrm obs}} (t)$
can be calculated by integrating the stellar orbits until time $t$, and 
$\s_{{\rho} , {\mathrm err}} (t)$ can be propagated from present using 
Eqs.~\ref{epic5}), we can also derive the time $t_{\mathrm min}$ at which 
$\s_{{\rho} , {\mathrm int}} (t)$ was minimum.  In Figure~\ref{fscoc1} we show the evolution
of $\s_{{\rho} , {\mathrm obs}}$, $\s_{{\rho} , {\mathrm err}}$ and 
$\s_{{\rho} , {\mathrm int}}$.  We observe a clear minimum in 
$\s_{{\rho} , {\mathrm int}}$ around $t_{\mathrm min} \approx 9 \cdot 10^6$~yr, 
which corresponds to B1 kinematic age. At the minimum spatial concentration 
we find the intrinsic spatial dispersions are $\s_{\xi^\prime}$~=~20~pc, 
$\s_{\eta^\prime}$~=~22~pc, and $\s_{\zeta^\prime}$~=~15~pc, smaller than their
current values of $\s_{\xi^\prime}$~=~30~pc, $\s_{\eta^\prime}$~=~33~pc, and
$\s_{\zeta^\prime}$~=~16~pc. Also, their intrinsic velocity dispersions were
smaller at $t_{\mathrm min}$, and $\la$~2~\kms\ in each component. 

  There is a clear discrepancy between the kinematic and the ``photometric" 
age of B1. This could be due to several reasons. First of all,
most of the stars in B1 are massive, and their atmospheres are probably 
rotating very quickly. Thus, the observed photometric colors are affected by 
this rotation, and so are the derived atmospheric parameters. The photometric 
ages, determined from these atmospheric parameters and stellar evolutionary 
models \cite{asiain0}, are systematically overestimated because of this
effect. In order to evaluate this effect, we have corrected the observed
photometric colors for rotation by considering a constant angle of inclination
between the line of sight and the rotation angle ($i = 60^\circ$), and a constant
atmospheric angular velocity ($\omega = 0.6 \; \omega_c$, where $\omega_c$ is 
the critical angular velocity). In this way we obtain a mean age $\approx 1.1 \cdot
10^7$~yr, and even smaller values for $\omega > 0.6 \; \omega_c$. More details on 
the correction for atmospheric rotation can be found in Figueras \& 
Blasi~\cite*{blasi98}. Second, before applying our method to
detect moving groups (Paper~I) we eliminated all stars with relative 
errors in age bigger than 100~\%, which slightly bias the age of young groups 
to larger values.

 In a recent study de~Zeeuw et al.~\cite*{dezeeuw99} carried out a 
census of nearby OB~associations using HIPPARCOS astrometric measurements 
and a procedure that combines both a convergent point method and a method 
that uses parallaxes in addition to positions and proper motions. Their
sample of neighbouring stars is much larger than ours, since radial 
velocities and Str\"omgren photometry are needed for our analysis. They found
521 stars in the Sco-Cen association, of which only 65 were in our 
sample. From these stars we have determined the mean velocity components of each 
association, which are in 
close agreement to those of B1 (Table~\ref{tassoc1}). However, their dispersions 
and ages are quite a bit larger than expected for a typical association. A 
deeper analysis revealed to us the presence of some stars in de~Zeeuw et 
al.'s~\cite*{dezeeuw99} associations whose peculiar velocities are responsible 
for their large 
velocity dispersion. Consistently, the ages of these stars are also peculiar
($\tau \approx 1-5 \cdot 10^8$~yr). Removing these stars and a few others with 
anomalously large ages\footnote{Eliminated stars: HIP79599 and HIP80126 in US; 
HIP70690,
HIP73147, HIP80208 and HIP83693 in UCL; HIP60379, HIP62703 and HIP64933 in LCC},
all of which probably belong to the field, we find a more reasonable 
kinematic properties and ages for these associations (Table~\ref{tassoc2}).

\begin{table}[htb]
\caption[]{Number of members ($\cal N$), spatial velocity components and ages 
(along with their dispersions) of the Sco-Cen associations 
found by de Zeeuw et al. \cite*{dezeeuw99}, determined from those members 
present in our sample}\label{tassoc1}
\begin{center}
\begin{tabular}{cccccc}
\hline
Assoc. & $\cal N$        & U               &  V               &  W               & $\tau$ \\
 &        & [km s$^{-1}$]   &  [km s$^{-1}$]   &  [km s$^{-1}$]   & [10$^8$~yr]  \\
\hline
  US &   14            & -3.5$_{(4.3)}$ &  -15.7$_{(2.1)}$   &  -6.4$_{(1.8)}$    & 0.4$_{(0.5)}$  \\
  UCL &   32            & -3.7$_{(7.0)}$ &  -19.7$_{(4.1)}$   &  -4.9$_{(2.1)}$    & 0.7$_{(1.0)}$ \\
  LCC &   19            & -6.8$_{(4.7)}$ &  -18.5$_{(6.5)}$   &  -6.4$_{(1.7)}$    & 0.9$_{(1.2)}$ \\
\hline
    TOTAL & 65            & -4.6$_{(6.1)}$ &  -18.5$_{(4.9)}$   &  -5.7$_{(2.1)}$    & 0.7$_{(1.0)}$ \\
\hline
\end{tabular}
\end{center}
\end{table}

\begin{table}[htb]
\caption[]{Same as Table~\ref{tassoc1}, now removing a few stars with peculiar
velocity components and ages} \label{tassoc2}
\begin{center}
\begin{tabular}{cccccc}
\hline
Assoc. & $\cal N$        & U               &  V               &  W               & $\tau$ \\
 &        & [km s$^{-1}$]   &  [km s$^{-1}$]   &  [km s$^{-1}$]   & [10$^8$~yr]  \\
\hline
  US &   12            & -3.8$_{(4.0)}$ &  -15.5$_{(2.1)}$   &  -6.6$_{(1.8)}$    & 0.2$_{(0.1)}$  \\
  UCL &   28            & -1.7$_{(4.4)}$ &  -20.4$_{(3.6)}$   &  -4.4$_{(1.5)}$    & 0.4$_{(0.3)}$ \\
  LCC &   16            & -5.1$_{(2.7)}$ &  -20.6$_{(4.0)}$   &  -6.0$_{(1.6)}$    & 0.4$_{(0.3)}$ \\
\hline
    TOTAL & 56            & -3.1$_{(4.2)}$ &  -19.4$_{(4.0)}$   &  -5.3$_{(1.9)}$    & 0.4$_{(0.3)}$ \\
\hline
\end{tabular}
\end{center}
\end{table}

\subsection{B2 group}\label{ssb2}

\begin{figure}
      \mbox{}
      \vspace{7.0cm}
      \includegraphics{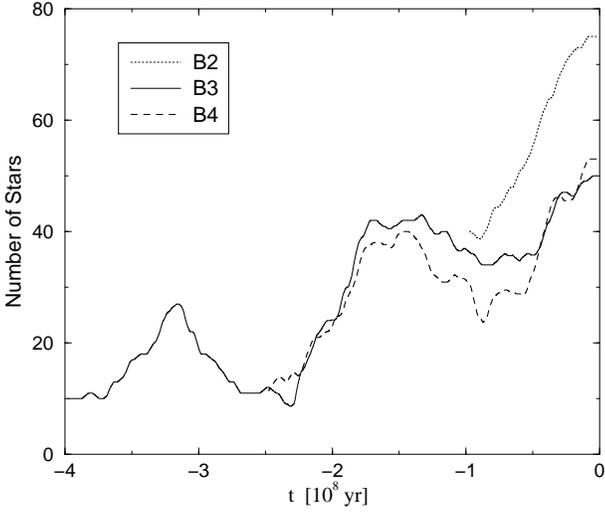}
      \caption[]{Number of MG stars placed around their nucleus (at a distance
		 $\le$~300~pc) as a function of time, for MGs B2, B3 and B4}
      \label{fconcen}
   \end{figure}

\begin{figure}
      \mbox{}
      \vspace{13cm}
      \includegraphics{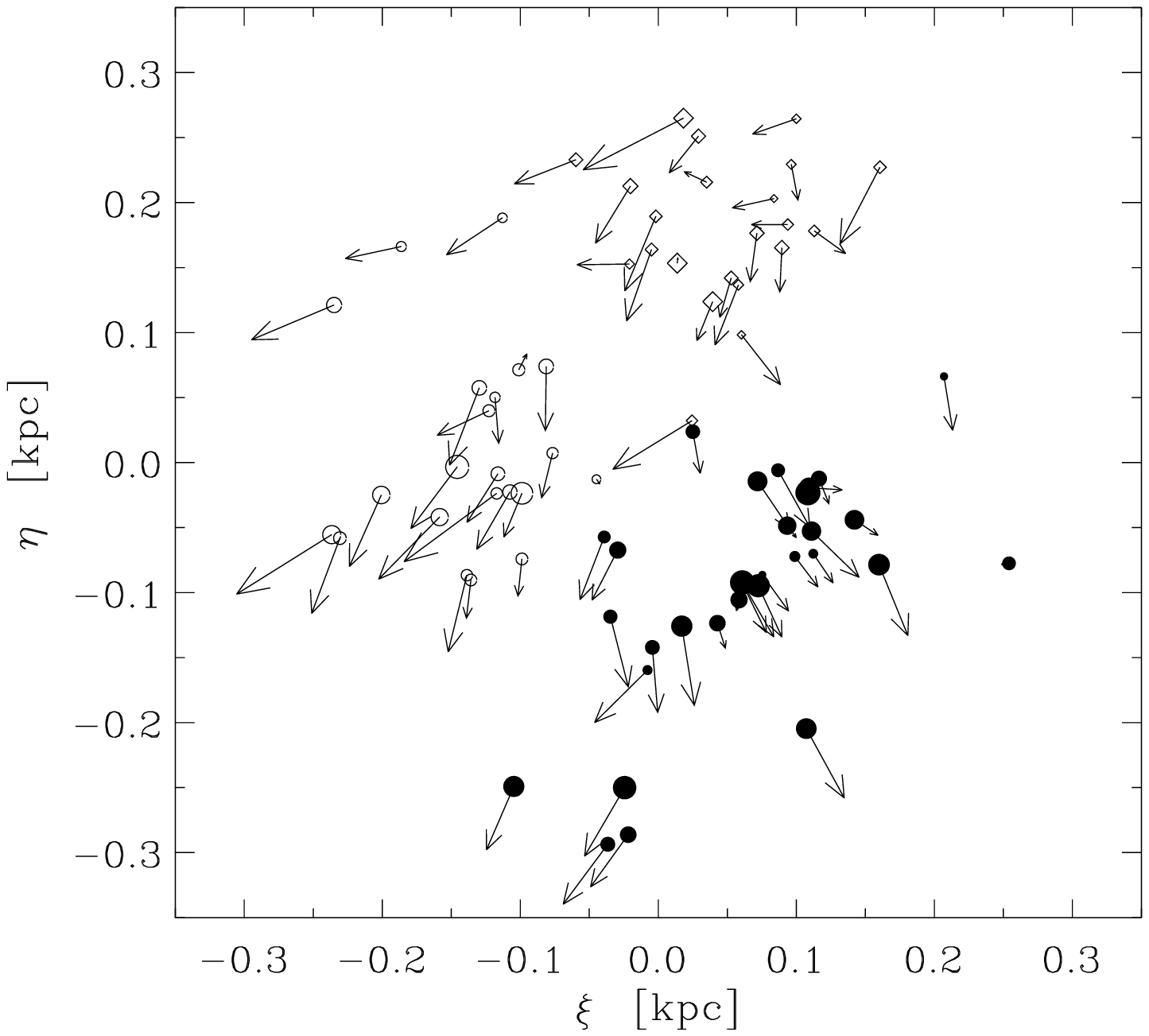}
      \includegraphics{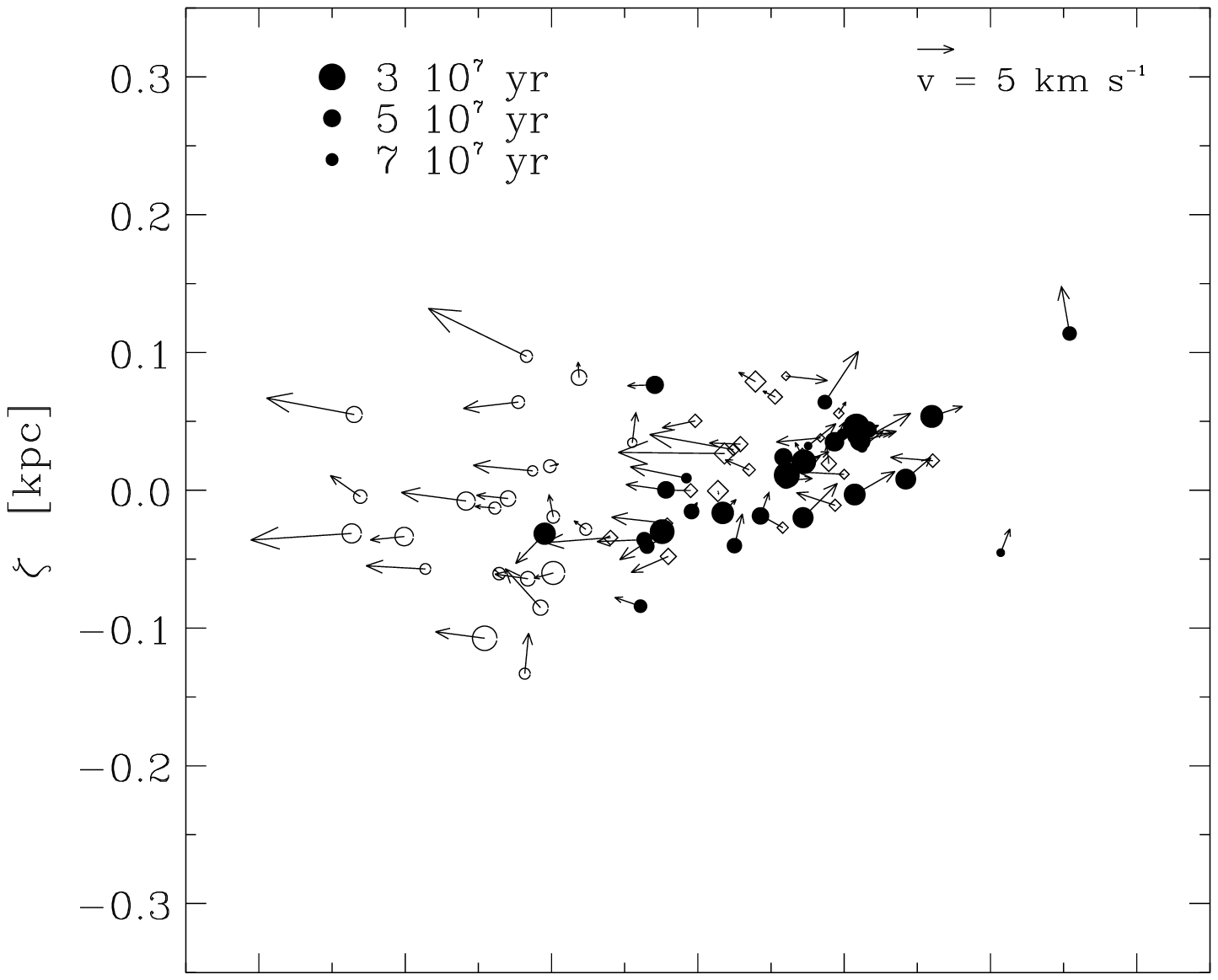}
      \caption[]{Position of B2 stars at present and their velocities 
                 referred to the LSR and corrected for galactic differential
                 rotation. Different symbols are used for three galactic
		 longitude ranges. The size of the symbols is inversely 
		 proportional to the age of the stars they represent}
      \label{fB2flet}
   \end{figure}

 Even though B2 is quite a young group, the propagation of age uncertainties 
over time prevents us from determining the group's kinematic age by means of the 
procedure developed for the B1 group. 
Instead, we have determined the trajectories of each star in B2, and
that of the MG's nucleus itself. The spatial concentration of these stars is 
propagated back in time by counting the number of stars found in 300~pc around
the MG nucleus (Fig.~\ref{fconcen}). We do not find any maximum in spatial 
concentration during the last $10^8$~yr. To better understand the structure of this 
group we plot in Fig.~\ref{fB2flet} the position of these stars and their 
velocity components on the galactic and meridional plane, referred to the LSR 
and corrected for galactic differential rotation. We observe that it is actually 
composed of several spatial stellar {\em clumps}, each of which has its own
kinematic behaviour, although certain degree of mixture is also evident. In 
particular, one of these clumps (most of the filled circles in 
Fig.~\ref{fB2flet}) perfectly overlaps with the Sco-Cen 
association mentioned above (20 of these stars are classified as members of 
this association by de Zeeuw et al. 1999). This
clumps's velocity components, and kinematic age, are also very similar to 
those found for the B1 group. Once again, the photometric age of these stars 
is probably overestimated because of the high rotation velocity of their 
atmospheres\footnote{Our MG finding algorithm (Paper~I) was able to detect 47 of the 56 
Sco-Cen association members in our sample. However, the classification
procedure we applied could not group them into a single MG. The age of these 
very young stars, affected by the important uncertainties already mentioned, is 
probably causing this misclassification}. A second clump is coincident with 
the Cassiopeia-Taurus (Cas-Tau) 
association in both position and velocity spaces (most of the empty circles 
in Fig.~\ref{fB2flet}). There are two dominant streams among these stars. 
In the heliocentric system, corrected for galactic differential rotation, their 
velocities are (U,V)$\simeq$(-10,-18)~\kms\ and (-13,-23)~\kms\ respectively 
(they share a common W component $\sim -6$~\kms). Their ages are 
$5-6 \cdot 10^7$~yr. These values are in good agreement with those found by 
de Zeeuw et al.~\cite*{dezeeuw99} for Cas-Tau association, i.e. 
(U,V,W) = (-13.24,-19.69,-6.38)~\kms\ with an age of $5\cdot 10^7$~yr (actually, 
about half of them are classified as Cas-Tau members by these authors). Finally, 
we observe a less concentrated group of
stars at $\xi \approx 0.05$~kpc and $\eta \approx 0.2$~kpc, or 
$l \approx 90^\circ$ (diamonds in Fig.~\ref{fB2flet}). The kinematic 
properties and age of this clump are very similar to those of the Cas-Tau 
association, whereas its position quite well agrees with that of the recently 
discovered Cepheus~OB6 association \cite{hooger97}.

   Thus, B2 seems to be the superposition of several OB associations 
from the Gould Belt, which are mixing with each other in the process of 
disintegration. These stars were classified as belonging to the same MG in
Paper~I since they roughly share the same kinematics and age, a consequence of 
their being formed from the same material. 

\subsection{Older Pleiades subgroups}\label{ssolder}

 Groups B3 and B4 in Table~\ref{tmgfi1} are considerably older than 
B1 and B2, and therefore only few details on the conditions in which they were 
formed can be recovered. In particular, the uncertainties in $\eta^\prime$
due to observational error increase almost monotonically with time. Though this 
increase can be closely approximated by Eq.~\ref{erecta} for an 
axysimmetric galactic potential, when considering the terms 
$\Phi_\mathrm{Sp}$ and $\Phi_\mathrm{B}$ this approximation breaks
(it only works during the first $\approx 10^8$~yr).  To estimate
the effect of typical errors in both the position ($\approx$~10-15~pc) 
and velocity ($\approx$~2-3~\kms) components of moving groups on 
$\eta^\prime$ as a function of time, we have simulated a stellar group 
whose dispersions in the phase space equals those errors, then determined 
their orbits. We obtain a dispersion in $\eta^\prime$ due to these errors 
of $\approx$~500~pc after $\sim 3 \cdot 10^8$~yr. Moreover, we cannot 
know the way in which individual orbits have been perturbed due to the 
disc heating effect, producing additional spatial and velocity dispersions.

   \begin{figure}
      \mbox{}
      \vspace{7.2cm}
      \includegraphics{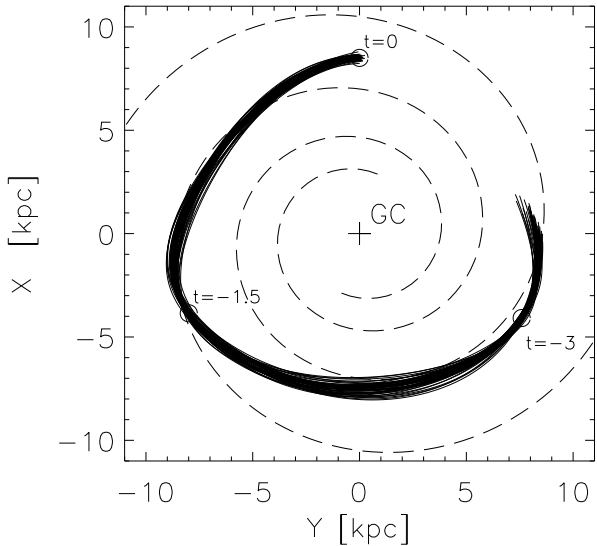}
      \caption[]{Stellar trajectories of the B3 members during the last    
                 3.5$\cdot 10^8$~yr. The position of the B3 nucleus at 
		 $t$ = 0, -1.5 and -3~$10^8$~yr (in units of $10^8$~yr) is
		 indicated with an open circle. The reference system is the
		 same as in Fig.~\ref{ftraj}}
      \label{fpos-ed1}
   \end{figure}

 As mentioned above, these older groups seem to have been born in the 
vicinity of the spiral arms. The position at birth of groups B3 and B4 are 
especially interesting: on the one hand, the trajectories followed by these
groups show a maximum galactocentric distance at the moment they were born,
which corresponds to a minimum kinematic energy (Figs.~\ref{ftraj} and 
\ref{fpos-ed1}); on the other hand, by 
determining the trajectories of the B3 members we observe two focusing points 
close to the B3 and B4 birthplaces (Fig.~\ref{fpos-ed1}). An identical result is
obtained when using B4 stars. Following Yuan~\cite*{yuan}, these points could 
be interpreted as the birthplace of B3 and B4. For spatially concentrated 
groups of stars with small velocity dispersions epicycle theory predicts 
(Sect.~\ref{sfoc}) the focusing phenomenon will be produced every  
$\Delta t = \displaystyle\frac{2 \pi}{\kappa}$. Since the Pleiades groups 
oscillate around a guiding 
center placed at $R \approx$~8.0-8.2~kpc, the corresponding epicycle frequency
is $\kappa \approx$~38-40~\kmskpc (determined from $\Phi_\mathrm{AS}$, 
Sect.~\ref{ssstetraj}), with $\Delta t \approx$~1.5-1.6~$\cdot 10^8$~yr. This
period $\Delta t$ is compatible with the ages of B3 and B4.  

 By following the same procedure used for the B2 group we study the 
spatial concentration of the older groups (Fig.~\ref{fconcen}). For B4 there 
is a maximum concentration at $t \approx -1.5~\cdot 10^8$, which corresponds 
to this MG's age, whereas we observe two peaks in concentration for B3 at 
$t \approx -1.5$ and $-3 ~\cdot 10^8$ respectively, the first one 
corresponding to the last focusing event, and the second one corresponding
to the average age of its members.

\section{Evolution of a Stellar Complex}\label{ssevsc}

  As defined by Efremov~\cite*{efre88}, Stellar Complexes (SC) are ``groupings of 
stars hundreds of parsecs in size and up to 10$^8$~yr in age, apparently uniting
stars born in the same gaseous complex". Associations and open clusters are the 
brighter and denser parts of these huge gaseous complexes. In other words, and 
according to Elmegreen \& Elmegreen \cite*{elme83}, the fundamental unit of star
formation is an HI-supercloud of $\sim 10^7$~M$_\odot$ which, during 
fragmentation into $\sim 10^5$-$10^6$~M$_\odot$ giant molecular clouds, gives 
birth to a SC. From these giant molecular clouds clusters and associations are 
formed. Examples of SCs are the Gould Belt in our Galaxy, 30 Doradus in 
the LMC, and NGC~206 in Andromeda. 

 In this section we explore the possibility that such objects are the
progenitors of MGs. Since both  galactic 
differential rotation and disc heating effect tend to disrupt any unbound 
system of stars on a short timescale, the large number of stars born inside 
a SC -- and therefore roughly sharing the same kinematics and 
ages -- may ensure a high spatial concentration for a longer time, especially
at the focusing points. 

   \begin{figure*}
     \vspace{11cm}	
      \includegraphics{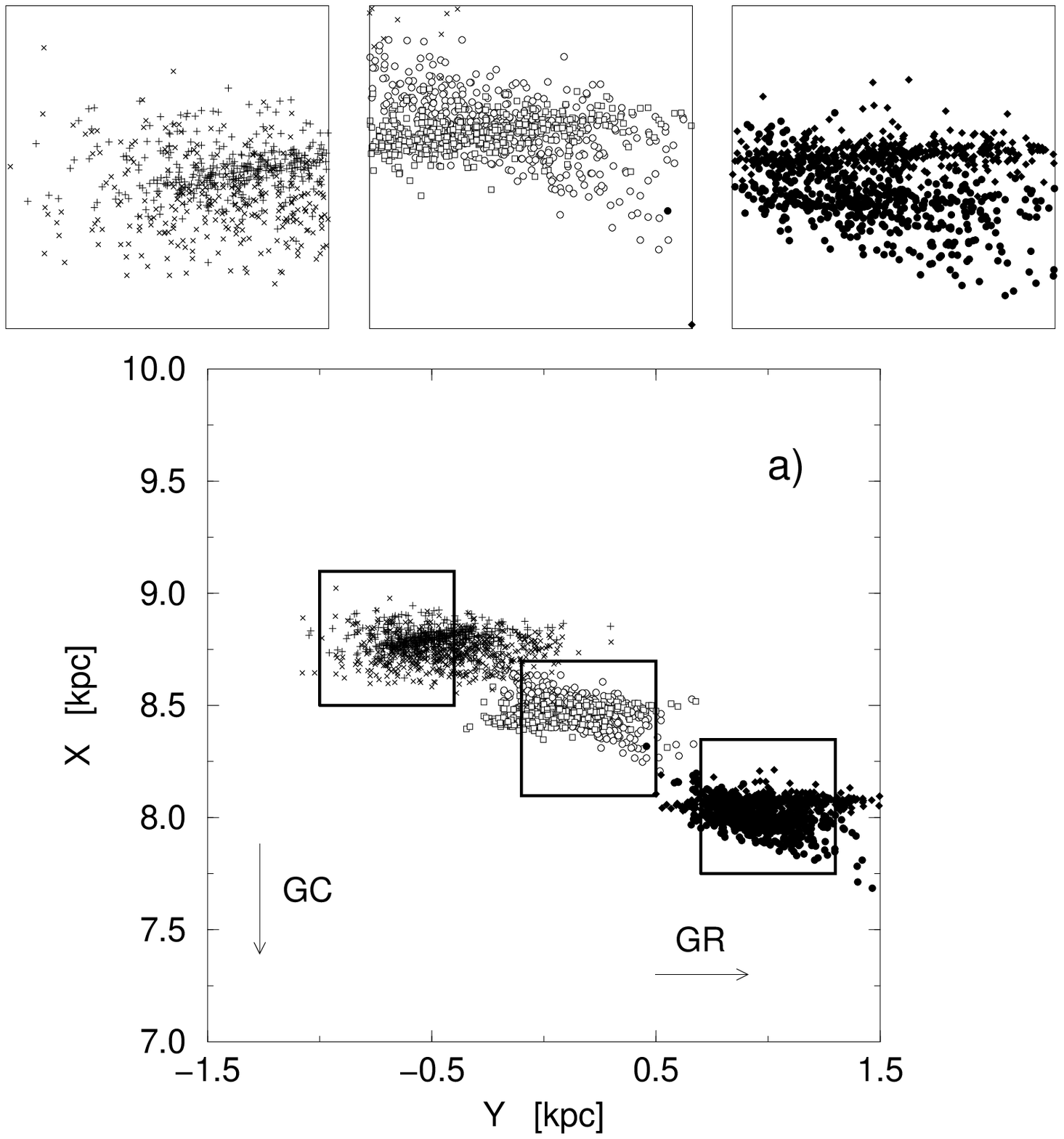}
      \includegraphics{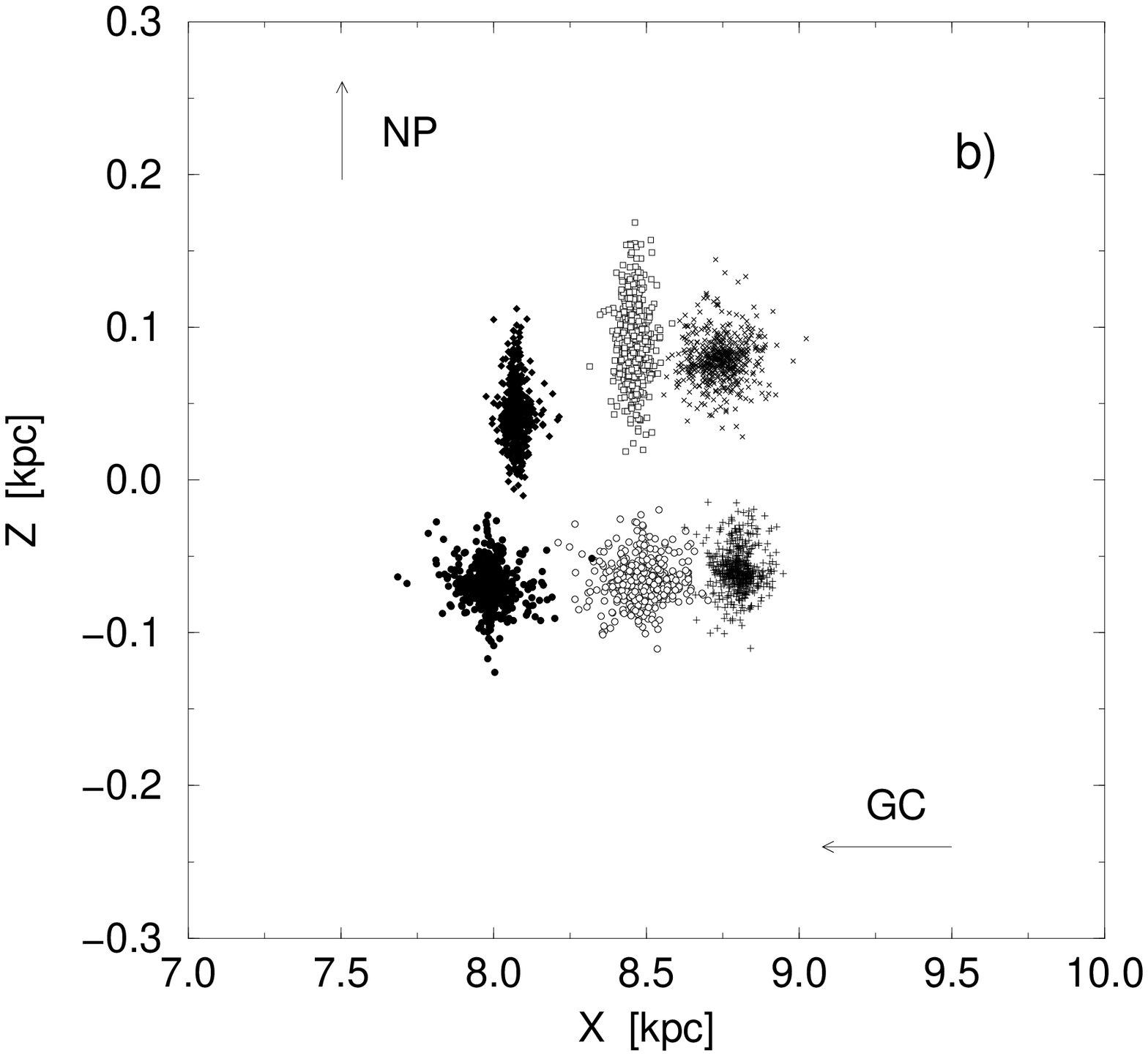}
      \caption[]{Position on the galactic ({\bf a}) and meridional ({\bf b}) 
		 planes of the simulated stellar complex at $t = 0$ (see text) 
		 when no disc heating effect is considered. Three square 
		 regions (600~pc/side) on the galactic plane have been 
		 enlarged to better distinguish the details. A different 
		 symbol is used for each group of stars coming from the same 
		 association. GC: galactic center; NP: north galactic pole; 
		 GR: galactic rotation} 
      \label{fste0} 
   \end{figure*}

 In order to analyse the possible link between SCs and 
MGs, we have generated a SC as a set of different unbound 
systems born at different epochs. Its main characteristics are taken from the 
literature and are described in what follows.  Although there are
no preferences on the galactic plane region where SCs are formed
(they can even be found in the interarm region, as suggested by the
observations of OB Associations in Andromeda, e.g.  Magnier et al. 1993),
we select the position and velocity of the B4 nucleus at birth 
($t = -1.5 \cdot 10^8$~yr), since in this way we can compare the results here 
obtained with observations on this real group. According to several 
authors' estimations \cite{efre88,elme85,efre94}, SCs are several hundreds of 
parsecs in size. The original shape of our simulated complex has been designed
as an ellipsoid whose equatorial plane is 250~pc in radius and lies on the 
galactic plane, and its vertical semi-axis is 70~pc. Six associations are born 
inside it at different epochs, randomly distributed inside this volume. Since 
the dispersion in age can be large when considering the SC as a whole 
\cite{efre88}, we impose the condition that one associations borns every  
10$^7$~yr. Hence, the first one is born at $t = -1.5 \cdot 10^8$~yr, the second one 
at $-1.4 \cdot 10^8$~yr, and so on.  The mean velocities of these associations
at birth follow a gaussian distribution around the B4 nucleus velocity, with an 
(isotropic) dispersion similar to the velocity dispersion inside a molecular 
cloud, i.e.  $\approx 5/\sqrt{3}$~\kms\ in each component \cite{stark89}.
Each of these associations is defined as follows: they contain 500~stars,
which are randomly distributed inside a sphere of 15~pc in diameter, an
intermediate value between the smallest associations (e.g. the Trapezium 
cluster) and the largest ones \cite{blaauw91,brown96}; and their internal 
velocity dispersion is 2~\kms\ in each velocity component, as observed in most 
of the closer OB~Associations \cite{blaauw91} and in molecular clouds 
\cite{scov90}; their internal dispersion in ages is 10~$^7$~yr \cite{efre88}. 
We also assume that all the simulated stars survive up to the time ($t$) when 
we study the system.

   \begin{figure*}
     \vspace{11cm}	
      \includegraphics{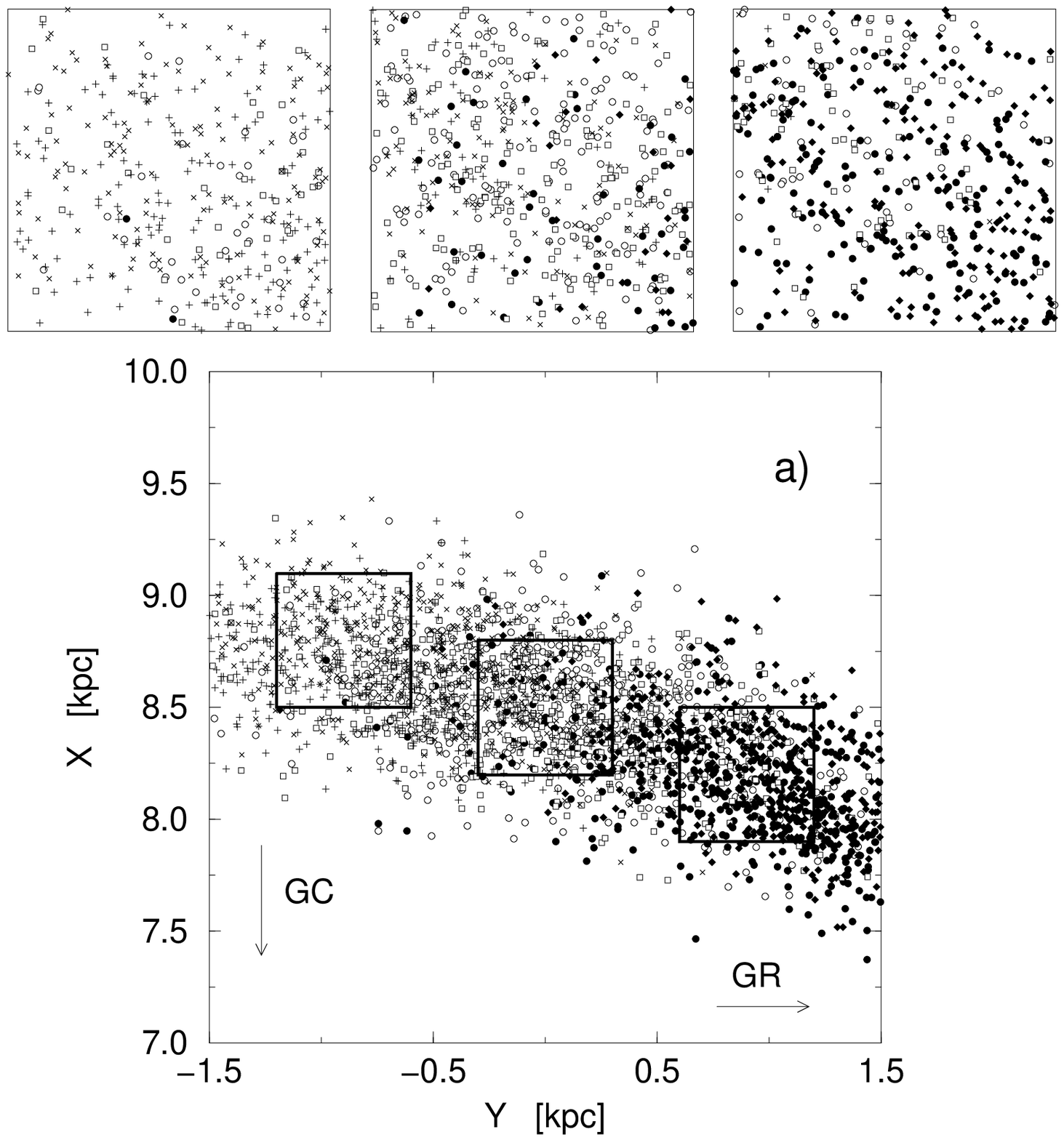}
      \includegraphics{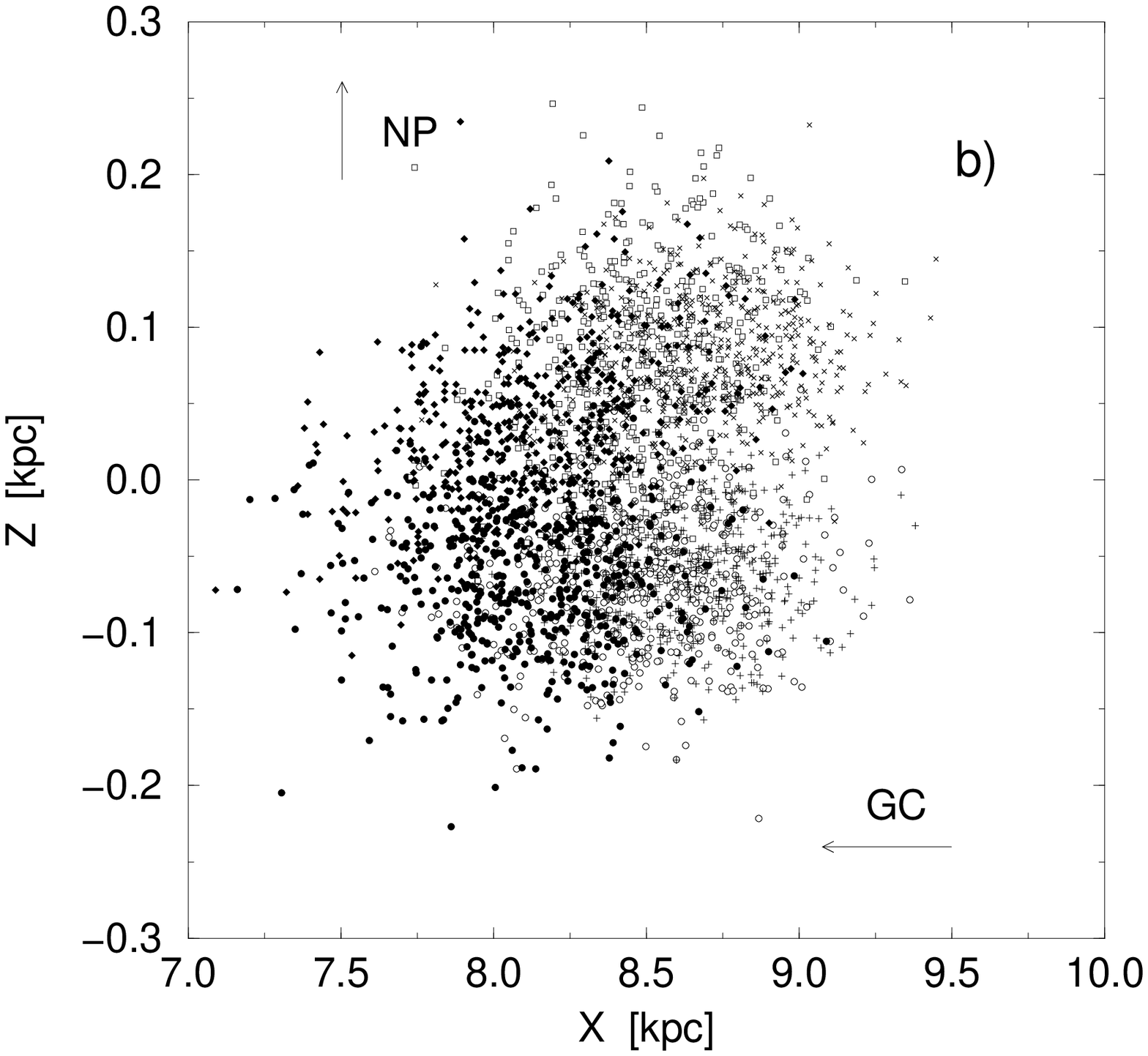}
      \caption[]{Same as Fig.~\ref{fste0} but considering now the effect
                 of the disc heating on the trajectory of the stars}
      \label{fste1} 
   \end{figure*}

 The orbit of each simulated star from the moment it was born up to the 
present ($t = 0$) is determined considering the galactic gravitational
potential ($\Phi_\mathrm{AS} +  \Phi_\mathrm{Sp} + \Phi_\mathrm{B}$), without 
taking into account the disc heating effect.
Their current spatial distribution is shown in Fig.~\ref{fste0}, where 
different symbols are used for the stellar 
``groups" coming from each simulated association.  According to our simulation, 
these groups are still concentrated in space, although their distribution is, 
at present, far from being isotropic due to the differential rotation. Each of 
them occupies an ellipsoidal volume $\approx$~150-250~pc wide, 500~pc long, and 
70~pc height, and are perfectly separated in space (four of them can be 
observed at different positions inside a 300~pc radius sphere around the Sun). 
The dispersion in each velocity component of any of the groups is
still $\approx$~2~\kms, as expected from Eqs.~\ref{epic5}. Although
the mean velocity of these stars is very similar to that of B4,
the dispersions in both velocity and space are too small when compared to
dispersions in detected MGs that are a few 10$^8$~yr old (Table~\ref{tmgfi1}). 
Thus, the spiral arms and central bar perturbations to the galactic 
gravitational potential cannot account for the observational MGs' velocity 
dispersions.

Results drastically change when considering the effect of constant heating  
on each individual star's trajectory, as described in Sect.~\ref{ssdh}. To
determine these trajectories we use only the axisymmetric part of the galactic 
potential, since the asymmetric perturbations to this potential are already
included in the treatment of the heating. The distribution of the simulated 
stars at present is shown in 
Fig.~\ref{fste1}. The members of the different groups are now completely 
mixed with each other. The total SC forms a long structure 
$\sim$~3~kpc long and $\sim$~1~kpc wide. Since in this simulation the Sun 
has been considered to be in the very center of the SC at 
$t = 0$, the members of this complex can be found up to some 500~pc from us in 
the radial direction, and much further in the tangential direction.
By enlarging a square region (600~pc/side) around the Sun we observe in 
Fig.~\ref{fste1} that, although the members of two groups dominate in the 
solar neighbourhood, all the other evolved associations are also represented. 
Again, the 
mean velocity components of B4 are recovered when considering only the
(636) stars closer than 300~pc from the Sun. However, as expected, the 
dispersions in velocities are now higher than in the former simulation -- the
total velocity dispersion is $\approx$~8-10~\kms, depending on the region of the
evolved SC. Hence, a constant diffusion of the stellar velocities can 
perfectly account for B4 velocity dispersions. 

 In order to study the properties of younger MGs, we analyse the general 
properties of the complex at its earlier stages, only $5 \cdot 10^7$~yr 
after the first association was born. At that moment 
($t = -1 \cdot 10^8$~yr) some stars in the youngest simulated groups are not 
yet born, while some others are as young as B1 and B2 members. 
The associations are still very concentrated in the phase space, though some
merging can be observed, as happens with B2 group. Velocity dispersions inside 
each association clearly depend upon their ages, as expected. Thus, in the 
U-component the velocity dispersion varies from $\approx$ 2.7~\kms\ for youngest 
groups to $\approx$ 4.1~\kms\ for the oldest (in the other components the 
dispersions are smaller).

 Finally, it is interesting to take a look to the general properties of
a simulated SC at a larger $t$. The mean properties of the simulated SC are now
taken from the B3 nucleus at birth ($t = -3 \cdot 10^8$~yr). At present ($t = 
0$~yr), assuming a constant heating of $\s_h = 1.45$~\kms\ every 
$\Delta t = 10^7$~yr, the 3000 simulated stars occupy a huge curved region, 
about 2~kpc wide and almost 10~kpc long. In the most dense parts of this region, 
$\sim$~200 stars can be found in a 300~pc radius sphere, whose total velocity 
dispersions are very high (12-14~\kms).  Stars with such large dispersions would 
be completely merged with field stars, so they could not be detected as MG 
members. A much smaller diffusion coefficient is needed to recover the B3's
velocity dispersions ($\s_h \approx 0.7$~\kms). 

 Thus, the disgregation of SCs by means of a constant heating mechanism is not 
able to account for the whole set of Pleiades MG substructures. A diffusion 
coefficient that depends on the stellar peculiar velocity and/or on the time 
(e.g. episodic diffusion) probably might explain the observed properties of 
these groups.  Moreover, the presence of some open clusters in the original SC 
could keep the stars together in phase space during longer periods. Some young 
open clusters, i.e. IC~2391, $\alpha$~Persei, Pleiades, etc, share roughly the same kinematics as the 
groups in Table~\ref{tmgfi1}, which favours this hypothesis. 

\section{Conclusions}

  In this paper we studied the disruption of unbound systems of stars as
a mechanism to understand the origin and evolution of moving groups. The 
epicycle theory was used to find an analytic expression for the time dependence 
of dispersion in both the stellar position and velocity coordinates.
We have obtained in this way a simple expression for the secular increase of 
the dispersion in the azimuthal galactic coordinate over time as a function of
the initial conditions. This increase in dispersion is, in fact, a direct 
consequence of the galactic
differential rotation. 

 In order to overtake the constraints of the epicycle theory, we concentrated 
our analysis on the determination of the
stellar trajectories using numerical integration of the equations of motion,
which provided us with an independent and more accurate estimation of the 
evolution of unbound systems. To perform this integration we used an analytic 
and axisymmetric galactic potential, along with spiral arm and central bar 
perturbations. This latter procedure allowed us to include random perturbations
that mimic the disc heating effect on stellar trajectories. 

 The trajectories followed by the members of the Pleiades moving 
group substructures found in Paper~I were used to compare 
the kinematic and photometric ages of these structures, and to establish
their position at birth. The youngest group, B1 (the Sco-Cen
association), was found to be most spatially concentrated some 
$9 \cdot 10^6$~yr ago, a value considerably smaller than its photometric age.
This is probably due to the effect of the high stellar atmospheric rotation of 
early type stars on the observed photometric colors. Groups B4 and B3 were born
at a time from the present equal to one and two times the epicycle period 
respectively, which means that they 
are spatially focused at present (probably that is the reason why we can 
observe them). At birth, these two groups were found to be close to the spiral 
arm structure. Concerning B2, a detailed analysis revealed to us that it is 
actually composed of several associations which are disintegrating at present.
As well,  
its averaged photometric age is probably overstimated, as in the case of B1.    

 We considered the evolution of a stellar complex under the influence of the
galactic gravitational potential as a mechanism to account for the main physical 
properties of moving groups. The high velocity dispersions of some of the 
Pleiades moving group substructures detected among B and A type stars could be 
recovered when the effect of the disc heating on the individual stellar 
trajectories was considered. At the same time, the disc heating can account for
the mixing of the stellar complex associations, although they 
are still clearly separated in phase space during the first 
tens of million years in the complexes' life (as observed for the groups B1 and 
B2). After only $1.5 \cdot 10^8$~yr from the birth of the stellar complex, the 
complex occupies an ellipsoidal area of $\sim 3 \times 1.5$~kpc$^2$, with its 
longest axis oriented in the direction of galactic rotation. If the Sun were 
presently located in the very center of this disrupted stellar complex, we 
would be able to find the complex's components up to very large distances on 
the galactic plane. 

 Thus, a constant heating mechanism (compatible with the observational heating
law) acting on the stars of a stellar complex can explain the velocity 
dispersions obtained for those Pleiades moving group substructures that 
are younger than $\sim 1.5 \cdot 10^8$. The properties of the older 
Pleiades substructures could probably be recovered by considering a diffusion 
coefficient depending on the velocity of the stars, and maybe on the time,
and/or 
the inclusion of open clusters in the simulated stellar complex.

%------------------%
% Acknowledgements %
%------------------%
\begin{acknowledgements}

  We would like to thank Dr. Ana E. G\'omez and Dr. J. Palou\u s for many 
stimulating discussions and suggestions. We are also indebted to F. Blasi, who
estimated the effect of the atmospheric rotation on the computed age of our
youngest stars. 

 This work was supported by CICYT under contract ESP97-1803, and by the PICS
programme (CIRIT). 

\end{acknowledgements}

%------------------%
% Appendices       %
%------------------%

\appendix

\section{Propagation of dispersions with time in the epicycle theory}
\label{aerrorep}

 Let us consider the vector 
$\vec{x}$ = ($\xi^\prime$,$\eta^\prime$,$\zeta^\prime$,$\dot{\xi}^\prime$,$\dot{\eta}^\prime$,$\dot{\zeta}^\prime$) containing the position and velocity of a 
star at a given time $t$. We are interested in determining the propagation of 
the observational dispersions over time, in the epicycle approximation frame
(Eqs.~\ref{epicm}, and their derivatives). If we define the vector 
$\vec{x}^{\circ}$~=~($\xi^\prime_\circ$,$\eta^\prime_\circ$,$\zeta^\prime_\circ$,$\dot{\xi}^\prime_\circ$,$\dot{\eta}^\prime_\circ$,$\dot{\zeta}^\prime_\circ$)
which contains the initial position and velocity of our star, then the initial
dispersions can be propagated with time as:
\begin{equation}\label{errep}
\s^2_{x_i} (t) = \displaystyle\sum_{j=1}^{6} \left(
\displaystyle\frac{\partial x_i (t)}{\partial x_j^{\circ}} \right)^2_t
\s^2_{x_j^\circ}
\end{equation}
where $\s_{x_i^{\circ}}$ is the observed dispersion in the variable $x_i$, and 
$\s_{x_i}(t)$ the dispersion at $t$. The dependence of variables
$\vec{x}$ on the initial values $\vec{x}^{\circ}$ is given by 
Equations~\ref{epic2}. If the coefficients of epicycle equations are 
included in a vector 
$\vec{c}$~=~($\xi^\prime_a$,$\xi^\prime_b$,$\eta^\prime_a$,$\zeta^{\prime}$,$\phi$,$\psi$)
then the matrix ${\cal A} = (a^i_j) \equiv 
\left( \displaystyle\frac{\partial x_i}{\partial x_j^{\circ}} \right)$ in
Equation~\ref{errep} can be determined as the matrix product 
${\cal A}$~=~${\cal B} {\cal F}$, where the matrix 
${\cal B} = (b^i_j) \equiv \left( \displaystyle\frac{\partial x_i}{\partial c_j} \right)$ and
${\cal F} = (f^i_j) \equiv \left( \displaystyle\frac{\partial c_i}{\partial x_j^{\circ}} \right)$ ($i, j = 1, \ldots, 6$). 
By partially differentiating 
Eqs.~\ref{epicm} we can determine the matrix~${\cal B}$
(Table~\ref{tmatb}), whereas matrix  ${\cal F}$ is determined by just 
differentiating
Eqs.~\ref{epic2} (Table~\ref{tmatf}). Now, from matrices ${\cal B}$ and 
${\cal F}$ we can determine the matrix~$\cal A$, given in Table~\ref{tmata}.
From matrix~$\cal A$, and by means of the relationship~\ref{errep}, the propagated
dispersions can be expressed as:

\begin{table*}
    \caption[]{Elements of the matrix ${\cal B} \equiv \left( 
\displaystyle\frac{\partial x_i}{\partial c_j} \right)$ (see text) }
  \begin{center}
  \begin{tabular}{l|cccccc}
      \hline
 $c_j$      & \multicolumn{6}{c}{$x_i$} \\
 & & & & & & \\
  & $\xi^\prime$ & $\eta^\prime$ & $\zeta^\prime$ & $\dot{\xi}^\prime$ & 
          $\dot{\eta}^\prime$ & $\dot{\zeta}^\prime$  \\
  \hline
$\xi^\prime_a$ &  1 & $\cos (\phi + \kappa t)$ & 0 & 0 & $-\xi^\prime_b \sin (\kappa t + \phi)$  & 0 \\ 
$\xi^\prime_b$ & -2  A $t$ &  $- \displaystyle\frac{2 \omega\ci}{\kappa} \sin(\kappa t + \phi)$ & 1 & 0 &  $- \displaystyle\frac{2 \omega\ci \xi^\prime_b}{\kappa} \cos(\kappa t + \phi)$ & 0 \\
$\eta^\prime_a$ & 0 & 0 & 0 & $\cos(\nu t + \psi)$ & 0 & $-\zeta^\prime_a \sin(\nu t + \psi)$ \\
$\zeta^{\prime}$ & 0 & $-\kappa \sin (\kappa t + \phi)$ & 0 & 0 & $-\xi^\prime_b \kappa \cos (\kappa t + \phi)$ & 0 \\
$\phi$ & -2 A & $- 2 \omega\ci  \cos (\kappa t + \phi)$ & 0 & 0 & $2 \omega\ci \xi^\prime_b \sin (\kappa t + \phi)$ & 0 \\
$\psi$ & 0 & 0 & 0 & $-\nu \sin(\nu t + \psi)$ & 0 & $-\nu \zeta^\prime_a \cos (\nu t + \psi)$ \\
      \hline
  \end{tabular}
  \end{center}
 \label{tmatb}
\end{table*}

\begin{table*}
    \caption[]{Elements of the matrix ${\cal F} \equiv 
\left( \displaystyle\frac{\partial c_i}{\partial x_j^{\circ}} \right)$ 
(see text)}
  \begin{center}
  \begin{tabular}{l|cccccc}
      \hline
 $x_j^\circ$      & \multicolumn{6}{c}{$c_i$} \\
 & & & & & & \\
 & $\xi^\prime_a$  & $\xi^\prime_b$ & $\eta^\prime_a$  & $\zeta^{\prime}$ & $\phi$ & $\psi$ \\
  \hline
$\xi^\prime_\circ$ & $- \displaystyle\frac{\omega\ci}{\mbox{B}}$ & 0 & 0 & 0 & $- \displaystyle\frac{1}{2 \mbox{B}}$ & 0 \\
$\eta^\prime_\circ$ & $\displaystyle\frac{\mbox{A} \cos(\phi)}{\mbox{B}}$ & 0 & 0 & $-\displaystyle\frac{\sin(\phi)}{\kappa}$ & $\displaystyle\frac{\cos(\phi)}{2 \mbox{B}}$ & 0 \\
$\zeta^\prime_\circ$ & 0 & 1 & 0 & $-\displaystyle\frac{2 \omega\ci}{\kappa^2}$ & 0 & 0 \\
$\dot{\xi}^\prime_\circ$ & 0 & 0 & $\cos(\psi)$ & 0 & 0 & $-\displaystyle\frac{\sin(\psi)}{\nu}$ \\
$\dot{\eta}^\prime_\circ$ & $-\displaystyle\frac{\mbox{A} \sin(\phi)}{\xi^\prime_b \mbox{B}}$ & 0 & 0 & $-\displaystyle\frac{\cos(\phi)}{\xi^\prime_b \kappa}$ &  $-\displaystyle\frac{\sin(\phi)}{2 \xi^\prime_b \mbox{B}}$ & 0 \\
$\dot{\zeta}^\prime_\circ$ & 0 & 0 & $-\displaystyle\frac{\sin(\psi)}{\zeta^\prime_a}$ & 0 & 0 & $-\displaystyle\frac{\cos(\psi)}{\nu \zeta^\prime_a}$ \\
\hline
  \end{tabular}
  \end{center}
 \label{tmatf}
\end{table*}

\begin{eqnarray}\label{epic5}
\s^2_{\xi^\prime} & = & \left[ \displaystyle\frac{\mbox{A} \cos (\kappa t) - \omega\ci}{\mbox{B}} \right]^2 \s^2_{\xi^\prime_\circ} +
                        \left[ \displaystyle\frac{\sin (\kappa t)}{\kappa} \right]^2 \s^2_{\dot{\xi}^\prime_\circ} + \nonumber \\
      & & + \left[ \displaystyle\frac{\cos (\kappa t) - 1}{2 \mbox{B}} \right]^2 \s^2_{\dot{\eta}^\prime_\circ} \nonumber \\
\s^2_{\eta^\prime}& = & \left[ \displaystyle\frac{2 \omega\ci \mbox{A}}{\mbox{B}} \left( t - \displaystyle\frac{\sin (\kappa t)}{\kappa} \right) \right]^2 \s^2_{\xi^\prime_\circ} +
      \s^2_{\eta^\prime_\circ} + \nonumber \\
      & & + \left[ \displaystyle\frac{2 \omega\ci}{\kappa^2} \left( \cos (\kappa t) - 1  \right) \right]^2 \s^2_{\dot{\xi}^\prime_\circ} + \nonumber \\
      & & + \left[ \displaystyle\frac{1}{\mbox{B}} \left( \mbox{A} t - \displaystyle\frac{\omega\ci}{\kappa} \sin (\kappa t) \right) \right]^2 \s^2_{\dot{\eta}^\prime_\circ} \nonumber\\
\s^2_{\zeta^\prime}& = &  \cos^2 (\nu t) \s^2_{\zeta^\prime_\circ} + \left[ \displaystyle\frac{\sin (\nu t)}{\nu} \right]^2 \s^2_{\dot{\zeta}^\prime_\circ} \nonumber\\
\s^2_{\dot{\xi}^\prime} & = & \left[ \displaystyle\frac{\mbox{A} \kappa}{\mbox{B}} \sin (\kappa t) \right]^2 \s^2_{\xi^\prime_\circ} +
      \cos^2 (\kappa t) \s^2_{\dot{\xi}^\prime_\circ} + \nonumber \\
    & & + \left[ \displaystyle\frac{\kappa}{2 \mbox{B}} \sin (\kappa t) \right]^2 \s^2_{\dot{\eta}^\prime_\circ} \nonumber\\
\s^2_{\dot{\eta}^\prime} & = & \left[ \displaystyle\frac{2 \mbox{A} \omega\ci}{\mbox{B}} \left( 1 - \cos (\kappa t) \right) \right]^2 \s^2_{\xi^\prime_\circ} +
      \left[ \displaystyle\frac{2 \omega\ci}{\kappa} \sin (\kappa t) \right]^2 \s^2_{\dot{\xi}^\prime_\circ} + \nonumber \\
      & & + \left[ \displaystyle\frac{1}{\mbox{B}} \left( \mbox{A}  -  \omega\ci \cos (\kappa t)   \right) \right]^2 \s^2_{\dot{\eta}^\prime_\circ} \nonumber\\
\s^2_{\dot{\zeta}^\prime} & = & \left[ \nu \sin (\nu t) \right]^2 \s^2_{\zeta^\prime_\circ} + \cos^2 (\nu t)\s^2_{\dot{\zeta}^\prime_\circ} \; \mbox{ . }
\end{eqnarray}

\begin{table*}
    \caption[]{Elements of the matrix ${\cal A} \equiv
\left( \displaystyle\frac{\partial x_i}{\partial x_j^{\circ}} \right)$
(see text)}
  \begin{center}
  \begin{tabular}{l|cccccc}
      \hline
 $x_j^\circ$      & \multicolumn{6}{c}{$x_i$} \\
 & & & & & & \\
  & $\xi^\prime$ & $\eta^\prime$ & $\zeta^\prime$ & $\dot{\xi}^\prime$ & 
          $\dot{\eta}^\prime$ & $\dot{\zeta}^\prime$  \\
  \hline
$\xi^\prime_\circ$ & $\displaystyle\frac{\mbox{A} \cos (\kappa t) - \omega\ci}{\mbox{B}}$ & 0 & 0 & $\displaystyle\frac{\sin (\kappa t)}{\kappa}$ & $\displaystyle\frac{\cos (\kappa t) - 1}{2 \mbox{B}}$ & 0 \\
$\eta^\prime_\circ$ & $\displaystyle\frac{2 \omega\ci \mbox{A}}{\mbox{B}} \left[ t - \displaystyle\frac{\sin (\kappa t)}{\kappa} \right]$ & 1 & 0 & $\displaystyle\frac{2 \omega\ci}{\kappa^2} \left[ \cos(\kappa t) - 1 \right]$ & $\displaystyle\frac{1}{\mbox{B}} \left[ \mbox{A} t - \displaystyle\frac{\omega\ci}{\kappa} \sin (\kappa t) \right]$ & 0 \\
$\zeta^\prime_\circ$ & 0 & 0 & $\cos (\nu t)$ & 0 & 0 & $\displaystyle\frac{\sin (\nu t)}{\nu}$ \\
$\dot{\xi}^\prime_\circ$ & $\displaystyle\frac{\mbox{A} \kappa}{\mbox{B}} \sin (\kappa t)$ & 0 & 0 & $\cos (\kappa t)$ & $\displaystyle\frac{\kappa}{2 \mbox{B}} \sin (\kappa t)$ & 0 \\
$\dot{\eta}^\prime_\circ$ & $\displaystyle\frac{2 \omega\ci \mbox{A}}{\mbox{B}} \left[ 1 - \cos (\kappa t) \right]$ & 0 & 0 & $\displaystyle\frac{2 \omega\ci}{\kappa} \sin (\kappa t)$ & $\displaystyle\frac{1}{\mbox{B}} \left[ \mbox{A} - \omega\ci \cos (\kappa t) \right]$ & 0 \\
$\dot{\zeta}^\prime_\circ$ & 0 & 0 & $\nu \sin (\nu t)$ & 0 & 0 & $\cos (\nu t)$ \\
\hline
  \end{tabular}
  \end{center}
 \label{tmata}
\end{table*}

From these equations we observe that all the dispersions in variables 
$\vec{x}$ oscillate around a constant value, except $\s^2_{\eta^\prime}$, 
whose averaged value increases with time. For small values of $t$ 
Eqs.~\ref{epic5} can be approximated by:
\begin{eqnarray*}
\s_{\xi^\prime} & \approx & \s_{\xi^\prime_\circ} \cr
\s^2_{\eta^\prime}& \approx & \s^2_{\eta^\prime_\circ} + t^2\ \s^2_{\dot{\eta}^\prime_\circ} \cr
\s_{\zeta^\prime}& \approx &  \s_{\zeta^\prime_\circ} \cr
\s_{\dot{\xi}^\prime} & \approx & \s_{\dot{\xi}^\prime_\circ} \cr
\s_{\dot{\eta}^\prime} & \approx &  \s_{\dot{\eta}^\prime_\circ} \cr
\s_{\dot{\zeta}^\prime} & \approx & \s_{\dot{\zeta}^\prime_\circ} \; \mbox{ . }
\end{eqnarray*} 

%------------%
% References %
%------------%

\end{document}